\documentclass[preprint,aps,eqsecnum]{revtex4}

\usepackage{graphicx}% Include figure files
\begin{document}
%\small
%\large
%\preprint{API/123-QED}
%\title
%{Stochastic Loewner Evolution}
%\author{Hans C. Fogedby}
%\email{fogedby@phys.au.dk}
%\affiliation {Department of Physics and
%Astronomy,
%University of Aarhus DK-8000, Aarhus C, Denmark\\
%and
%\\
%Niels Bohr Institute, Blegdamsvej 17, DK-2100, Copenhagen {\O},
%Denmark }
{\bf{\large STOCHASTIC LOEWNER EVOLUTION}

Linking universality, criticality and conformal invariance \\ in
complex systems }

Hans C. Fogedby

Aarhus University, Aarhus, Denmark\newline and
\newline Niels Bohr Institute, Copenhagen, Denmark

{\bf ARTICLE OUTLINE}

I. ~~~~ Introduction
\newline
$~~~~~~~~~~\mbox{A. General Remarks}$
\newline
$~~~~~~~~~~\mbox{B. Scaling Ideas}$
\newline
$~~~~~~~~~~\mbox{C. Scaling in Equilibrium}$
\newline
$~~~~~~~~~~\mbox{D. Stochastic Loewner Evolution}$
\newline
$~~~~~~~~~~\mbox{E. Outline}$
\newline
II. ~~~ Scaling
\newline
$~~~~~~~~~~\mbox{A. Random Walk}$
\newline
$~~~~~~~~~~\mbox{B. Percolation}$
\newline
$~~~~~~~~~~\mbox{C. Ising Model}$
\newline
$~~~~~~~~~~\mbox{D. Critical curves - Exploration}$
\newline
$~~~~~~~~~~\mbox{E. Distributions - Markov properties - Measures}$
\newline
III. ~~ Conformal Invariance
\newline
$~~~~~~~~~~\mbox{A. Conformal Maps}$
\newline
$~~~~~~~~~~\mbox{B. Measures - Conformal Invariance}$
\newline
IV. ~~ Loewner Evolution
\newline
$~~~~~~~~~~\mbox{A. Growing Stick}$
\newline
$~~~~~~~~~~\mbox{B. Loewner Equation}$
\newline
$~~~~~~~~~~\mbox{C. Exact Solutions}$
\newline
V. ~~~ Stochastic Loewner Evolution
\newline
$~~~~~~~~~~\mbox{A. Schramm's Theorem}$
\newline
$~~~~~~~~~~\mbox{B. SLE Properties}$
\newline
$~~~~~~~~~~\mbox{C. Curves and Hulls - Bessel Process}$
\newline
$~~~~~~~~~~\mbox{D. Fractal Dimension}$
\newline
VI. ~~ Results and Discussion
\newline
$~~~~~~~~~~\mbox{A. Phase transitions - Locality - Restriction -
Duality}$
\newline
$~~~~~~~~~~\mbox{B. Loop Erased Random Walk}$
\newline
$~~~~~~~~~~\mbox{C. Self Avoiding Random Walk}$
\newline
$~~~~~~~~~~\mbox{D. Percolation}$
\newline
$~~~~~~~~~~\mbox{E. Ising model - O(n) Models}$
\newline
$~~~~~~~~~~\mbox{F. SLE - Conformal Field Theory}$
\newline
$~~~~~~~~~~\mbox{G. SLE - 2D Turbulence}$
\newline
$~~~~~~~~~~\mbox{H. SLE - 2D Spin glass}$
\newline
$~~~~~~~~~~\mbox{I. Further remarks}$
\newline
VII. ~ Future Directions
\newline
VIII. ~Bibliography

\newpage
\section{Introduction}

Stochastic Loewner evolution also called Schramm Loewner evolution
(abbreviated, SLE) is a rigorous tool in mathematics and statistical
physics for generating and studying scale invariant or fractal
random curves in two dimensions (2D). The method is based on the
older deterministic Loewner evolution introduced by Karl L\"owner
[76], who demonstrated that an arbitrary curve not crossing itself
can be generated by a real function by means of a conformal
transformation. A real function defined in one spatial dimension
(1D) thus encodes a curve in 2D, in itself an intriguing result. In
2000 Oded Schramm [82] extended this method and demonstrated that
driving the Loewner evolution by a 1D Brownian motion, the curves in
the complex plane become scale invariant; the fractal dimension
turns out to be determined by the strength of the Brownian motion.

The one-parameter family of scale invariant curves generated by
SLE is conjectured and has in some cases been proven to represent
the continuum or scaling limit of a variety of interfaces and
cluster boundaries in lattice models in  statistical physics,
ranging from self-avoiding random walks to percolation cluster
boundaries, and Ising domain walls.

SLE operates in the 2D continuum where it generates extended scale
invariant objects. SLE delimits scaling universality classes by a
single parameter $\kappa$, the strength of the 1D Brownian drive,
yielding the fractal dimension $D$ of the scale invariant shapes
according to the relation $D=1+\kappa/8$. Moreover, SLE provides the
geometrical aspects of conformal field theory (CFT). The central
charge $c$, delimiting scaling universality classes in CFT, is thus
related to $\kappa$ by means of the expression
$c=(3\kappa-8)(6-\kappa)/2\kappa$.

Stochastic Loewner evolution derives its' importance from the fact
that it addresses the issue of extended random fractal shapes in 2D
by direct analysis in the continuum. It thus supplements and extends
earlier lattice results and also allows for the determination of new
scaling exponents. From the point of view of conformal field theory
based on the concept of a local field, operator expansions, and
correlations, the geometrical approach afforded by SLE, directly
addressing conformally invariant random shapes in the continuum,
represents a novel point of view of maybe far reaching consequences;
so far only explored in two dimensions.

The field of SLE has mainly been driven by mathematicians presenting
their results in long and difficult papers. There are, however,
presently several excellent reviews of SLE both addressing the
theoretical physics community [5,6,12,13,21,57] and the mathematical
community [24,35], see also a complete biography up to 2003 [57].
The purpose of this article is to present a heuristic and simple
discussion of some of the key aspects of SLE, for details and topics
left out we refer the reader to the reviews mentioned above.
However, in order provide the necessary background and set the stage
for SLE we begin with some general remarks on scaling in statistical
physics.

\subsection{General Remarks}

In statistical physics we study macroscopic systems composed of many
interacting components. In the limit of many degrees of freedom the
macroscopic behavior roughly falls in two categories. In the most
common case the macroscopic behavior is deterministic and governed
by phenomenological theories like for example thermodynamics and
hydrodynamics operating entirely on a macroscopic level. This
behavior is basically a result of the law of large numbers,
permitting an effective coarse graining and yielding for example a
macroscopic density or velocity field [14]. In the other case, the
macroscopic behavior is dominated by fluctuations and shows a random
behavior [14,31]. Typical cases are random walk and equilibrium
systems tuned to the critical point of a second order phase
transition. Other random cases are for example self-organized
critical systems purported to model earth quake dynamics, flicker
noise and turbulence in fluids [4,20].

The distinction between the deterministic and random cases of
macroscopic behavior is illustrated by the simple example of a
biased random walk described by the Langevin equation
$dx(t)/dt=v+\xi(t)$, $\langle\xi(t)\xi(0)\rangle\propto\delta(t)$.
Here $x(t)$ is a macroscopic variable sampling the statistically
independent microscopic steps $\xi(t)$; the velocity $v$ is an
imposed drift or bias. Averaging over the steps we obtain for the
deviation of $x$, $R=[\langle x^2\rangle]^{1/2}=[(vt)^2+t]^{1/2}$.
For large $t$ the deviation $R\sim\langle x\rangle=vt$ and the
mean value or deterministic part dominates the behavior, the
fluctuational or random part $R\sim t^{1/2}$ being subdominant.
Fine tuning the random walk to vanishing bias $v=0$ we have
$\langle x\rangle =0$ and $R\sim t^{1/2}$, i.e., the random
fluctuations control the phenomenon.

The study of complexity encompasses a broader field that statistical
physics and is concerned with the emergence of universal properties
on a mesoscopic or macroscopic scale in large interacting systems.
For example particle systems, networks in biology and sociology,
cellular automata, etc. The class of complex systems generally falls
in the category of random systems. The emergent properties are a
result of many interacting agents or degrees of freedom and can in
general not be directly inferred from the microscopic details. A
major issue is thus the understanding of generic emerging properties
of complex systems [22,28,34]. Here, however, the methods of
statistical physics is an indispensable tool in the study of
complexity.

The evolution of statistical physics, a branch of theoretical
physics, has occurred in steps and is driven both by the
introduction of new concepts and the concurrent development of new
mathematical methods and techniques. A well-known case is the long
standing problem of second order phase transitions or critical
phenomena which gave way to a deeper understanding in the sixties
and seventies and spun off the renormalization group techniques for
the determination of critical exponents and universality classes
[8,11,26,32,36].

\subsection{Scaling Ideas}

This brings us to the fundamental scaling ideas and techniques
developed particularly in the context of critical phenomena in
equilibrium systems and which now pervade a good part of theoretical
physics and, moreover, play an important role in the analysis of
complex systems in physical sciences [8,11,14]. Scaling is
synonymous with no scale in the sense that a system exhibiting scale
invariance is characterized by the absence of any particular scale
or unit. Scaling occurs both in the space and/or time behavior and
is typically characterized by power law dependencies controlled by
scaling exponents.

A classical case is random walk discussed above, characterized by
the Langevin equation $dx/dt=\xi(t)$,
$\langle\xi\xi\rangle\sim\delta(t)$ [31]. Here the mean square
displacement scales like $\langle x^2\rangle(t)\sim t^{2H}$, where
$H$ is the Hurst scaling exponent; for random walk $H=1/2$ [16,27].
Correspondingly, the power spectrum $P(\omega)=|x_\omega|^2$,
$x_\omega=\int dt~ x(t)\exp(i\omega t)$, scales like $P(\omega)\sim
\omega^{-1-2H}$, i.e., $P(\omega)\sim\omega^{-2}$ for random walk;
we note that the underlying reason for the universal scaling
behavior of random walk is the central limit theorem [18,31].

\subsection{Scaling in Equilibrium}

Scaling ideas and associated techniques first came to the forefront
in statistical physics in the context of second order phase
transitions or critical phenomena [8,11,14,26]. More specifically,
consider the usual Ising model defined on a lattice with a local
spin degrees of freedom, $\sigma_i=\pm 1$ at site $i$, subject to a
short range interaction $J$ favoring spin alignment. The Ising
Hamiltonian has the form $H=-J\sum_{\langle
ij\rangle}\sigma_i\sigma_j$, where $\langle ij\rangle$ indicates
nearest neighbor sites. The thermodynamic phases are characterized
by the order parameter $m=\langle\sigma_i\rangle$. Above one
dimension the Ising model exhibits a second order phase transition
at a finite critical temperature $T_c$. Above $T_c$ the model is in
a disordered paramagnetic state with $m=0$ and only microscopic
domain of ordered spins. Below $T_c$ the model favors a
ferromagnetic state with long range order and macroscopic domains of
ordered spins, corresponding to $m\neq 0$; at $T=0$ the model
assumes the ferromagnetic ground state configuration with totally
aligned spins. Regarding the spatial organization, the size of the
domains of ordered spins is characterized by the correlation length
$\xi(T)$. As we approach the critical point at $T_c$ the order
parameter vanishes $m\propto|T-T_c|^\beta$ with critical exponent
$\beta$, but more significantly, the correlation length $\xi(T)$
diverges like $\xi(T)\sim |T-T_c|^{-\nu}$ with scaling exponent
$\nu$. This indicates that the system is scale invariant at $T_c$.
Regarding the domains of ordered spins, the system is spatially
self-similar on scales from the microscopic lattice distance to the
system size; the system size diverging in the thermodynamic limit.
The scaling behavior at the critical point $T_c$ is an emergent
property in the sense that the scaling exponents $\beta$ and $\nu$
do not depend on microscopic details like the type of lattice,
strength of interaction, etc., but only on the dimension of the
system and the symmetry of the order parameter [30].

The diverging correlation length at the critical point is the
central observation which in the 60-ties and 70-ties gave rise to a
detailed understanding of critical phenomena, beginning with the
coarse graining block scheme proposed by Kadanoff [58] and
culminating with the development and application of field
theoretical renormalization group techniques by Wilson and others
[8,11,14,26,30,36].

For a diverging correlation length much larger than the lattice
distance the local spin $\sigma_i$ can be replaced by a
coarse-grained local field $\phi(r)$ and the Ising Hamiltonian $H$
by the Ginzburg-Landau functional $F=\int
d^dr[(\nabla\phi)^2+R\phi^2+ U\phi^4$], where the 'mass' term
$R\sim|T-T_c|$. Consequently, the universality class of Ising-type
models is described by a scalar field theory. The renormalization
group techniques basically quantify the Kadanoff block
construction in momentum space and extract scaling properties in
an expansion about the upper critical dimension $d=4$. To leading
order in $4-d$ one obtains $\beta=1/2-(4-d)/6$ and
$\nu=1/2+(4-d)/12$ (Wilson 1974). Alternatively, keeping the
correlation length $\xi$ fixed and letting the lattice distance
approach zero, we obtain at $T_c$ the so-called scaling limit or
continuum limit of the Ising model. The Ising spin $\sigma_i$
becomes a local field $\phi(r)$ and the weight of a configuration
is determined by $\exp(-F)$. Note that in order to implement the
scaling limit we must be at $T_c$. The two scenarios of a growing
correlation length for fixed lattice distance implementing the
Kadanoff construction and a fixed correlation length for a
vanishing lattice distance yielding a continuum field theory are
related by an overall scale transformation [36].

It is generally assumed that the global or nonlocal scale
invariance at the critical point in the continuum limit can be
extended to a local scale invariance including translation and
rotation, that is an angle-preserving conformal transformation.
This follows heuristically from an a local implementation of the
Kadanoff coarse-graining block construction and applies to lattice
models with short range interactions and discrete translational
and rotational invariance [9,11]. The resulting continuum theories
then fall in the category of conformal field theories (CFT).

In 2D the group of conformal transformations is particularly rich
since it corresponds to the class of analytical functions $w=f(z)$
mapping the complex plane $z$ to the complex plane $w$. The infinite
group structure imposes sufficient constraints on the structure of
conformal field theories in 2D that the scaling form of
correlations, e.g., $\langle\phi\phi\rangle(r)$, and in particular
the critical exponents can be determined explicitly [10,11]. On
finds $\beta=1/8$ and $\nu=1$ for the order parameter and
correlation length exponents, respectively, in accordance with
lattice theory results (Baxter 1982). Here we also mention the
Coulomb gas method for the determination of critical exponents [29].

It is a common feature of both renormalization group calculations
based on an expansion about a critical dimension and conformal
field theory in 2D that the local field $\phi(r)$ and its
correlations are the basic building blocks and that the critical
properties are encoded in their scaling form, yielding critical
exponents, scaling laws, scaling functions, etc. Notwithstanding
the fact that the seminal Kadanoff construction [58] was based on
a geometrical picture corresponding to coarse-graining the degrees
of freedom over larger and larger scales, keeping track of ordered
domains on all scales, the actual geometry of critical phenomena
such as the scaling properties of critical clusters was not
well-understood and seemed inaccessible in the continuum limit
within the context of local field theories. Whereas it is not
difficult to generate critical domain walls, interfaces, and
clusters for lattice models with appropriate boundary conditions
by means of standard Monte Carlo simulation techniques, the
continuum or scaling limit of critical shapes appeared until
recently, with a few exceptions, beyond present techniques.

\subsection{Stochastic Loewner Evolution}

Here stochastic Loewner evolution (SLE) represents a new insight in
2D critical phenomena with respect to a deeper understanding of
scale invariant curves, clusters, and shapes. Also, there appears to
be deep connections between SLE and conformal field theory.

SLE is an ingenious way of generating critical curves and shapes
in the 2D continuum using conformal transformations. Let us
mention a characteristic example. Consider an Ising model on a
lattice in a chosen domain. Imposing spin up on a continuous part
of the domain boundary and spin down on the remaining part of the
boundary, it follows that a specific domain wall or interface will
connect the two points on the boundary where a bond is broken. At
low temperature the bond energies dominate and the free energy is
lowest for a straight domain wall with few kinks. As we approach
the critical point entropy or fluctuations come strongly into play
and the domain wall meanders balancing energy and entropy. At the
critical point the system becomes scale invariant with a diverging
correlation length and likewise the domain wall becomes scale
invariant, i.e., it has kinks on all scales larger that the
lattice distance. In the continuum limit the Ising domain wall
becomes a random fractal curve with a particular fractal
dimension. Here SLE provides a direct analytical method in the
continuum to generate such a random curve and, moreover, provides
the fractal dimension in terms of the strength of the 1D random
walk driving the SLE evolution.

\subsection{Outline}

The outline of the present particle is as follows. In Section III
on scaling we introduce some of the basic models and concepts
necessary for a discussion of SLE: A) Random walk, B) Percolation,
C) Ising model, D) Critical curves and exploration, and E)
Distributions, Markov properties, and measures. In Section IV we
turn to the essential ingredient in SLE, namely, conformal
invariance: A) Conformal maps and B) Measures and conformal
invariance. Section V is devoted to deterministic or classical
Loewner evolution: A) Growing stick, B) Loewner equation and C)
Exact solutions. In Section VI, constituting the core part of this
article, we discuss stochastic Loewner evolution: A) Schramm's
theorem, B) SLE properties, C) Curves, hulls, and the Bessel
process and D) Fractal dimension. Section VII is devoted to
results and discussions: A) Phase transitions, locality,
restriction, and duality, B) Loop erased random walk, C)
Self-avoiding random walk D) Percolation, E) Ising model and O(n)
models, F) SLE and conformal field theory, G) Application to 2D
turbulence, H) Application to 2D spin glass and I) Further
remarks. Finally, in Section VIII we discuss future directions of
the field. Section IX contains a bibliography including books,
general reviews, and more technical papers.

\section{Scaling}

Stochastic Loewner evolution, has been applied to a host of lattice
models proved or conjectured to possess scaling limits. However, for
the present purpose we will focus on three lattice models: Random
walk, Percolation, and the Ising model.

\subsection{Random Walk}

Random walk is a simple and much studied random process [3,16,31].
Consider an unbiased random walk in the plane composed of $N$
steps, where the i-th step, $\vec\eta_i$, is random, isotropic and
uncorrelated, i.e., $\langle\vec\eta_i\rangle=0$ and
$\langle\eta_i^\alpha\eta_j^\beta\rangle\propto
\delta_{\alpha\beta}\delta_{ij}$. For the end-to-end distance we
have $\vec x=\sum_{i=1}^N\vec\eta_i$ and for the size
$R=[\langle\vec x^2\rangle]^{1/2}\sim N^{1/2}$. Assuming one step
pr unit time, $N\propto t$, we obtain $R\sim t^{1/2}$,
characteristic of diffusive motion. Introducing the fractal
dimension $D$ by the usual box counting procedure [16,27]
\begin{eqnarray}
N(R)\sim R^D, \label{frac1}
\end{eqnarray}
where $N$ is the number of boxes and $R$ the size of the object, and
covering the random walk we readily infer $D=2$, i.e., the
self-crossing random walk is plane-filling modulo the lattice
distance. Introducing the scaling exponent $\nu$ according to $R\sim
N^\nu$ we have for random walk $\nu=1/2$; note that $D=1/\nu$.

The scaling limit of unbiased random walk is Brownian motion (BM)
[3] and is obtained by scaling the step size $\eta$ down and the
number of steps $N$ up in such a manner that the size $R\sim
N^{1/2}\eta$ stays constant. The resulting BM path is a continuous
non-differentiable random curve with fractal dimension $D=2$. The
BM path is plane-filling and recurrent in 2D, i.e., it returns to
a given point with probability one. Focussing on one of the
independent cartesian components 1D BM, $B_t$, is also described
by the Langevin equation
\begin{eqnarray}
\frac{dB_t}{dt} = \eta_t,~~\langle\eta_t\eta_s\rangle =\delta(t-s),
\label{lan}
\end{eqnarray}
where $\eta_t$ is uncorrelated Gaussian white noise with a flat
power spectrum. Integrating Eq. (\ref{lan}) $B_t$ samples the step
$\eta_t$ and we find, assuming $B_0=0$, $B_t=\int_0^t
\eta_{t'}dt'$ from which we directly infer the fundamental
properties of BM, namely, independence and stationarity,
\begin{eqnarray}
&&B_{T+\Delta T}-B_T\approx B_{\Delta T},
~~~~~~~~~~~~~~~~~~~~\rm(stationarity) \label{sta}
\\
&&B_{\Delta T}, B_{\Delta T'}~~\text{indep. for} ~~\Delta T
\neq\Delta T',~~\rm{(independence)} \label{ind}
\end{eqnarray}
where $\approx$ indicates identical distributions. Moreover, the
correlations are given by
\begin{eqnarray}
\langle|B_t-B_s|^2\rangle = |t-s|. \label{corr}
\end{eqnarray}
and $B_t$ is distributed according to the normal (Gaussian)
distribution $P(B,t)=(2\pi t)^{-1/2}\exp(-B^2/2t)$.

Whereas 1D BM drives SLE, 2D BM is not itself generated by SLE since
the path is self-crossing on all scales; by construction SLE is
limited to the generation of non-crossing random curves. However,
variations of BM have played an important role in the development of
SLE. We mention here the scaling limit of loop erased random walk
(LERW) and self-avoiding random walk (SAW), to be discussed later.

\subsection{Percolation}

The phenomenon of percolation is relevant in the context of
clustering, diffusion, fractals, phase transitions and disordered
systems. As a result, percolation theory provides a theoretical and
statistical background to many physical and natural sciences [33].

Percolation is the simplest lattice model exhibiting a geometrical
phase transition.  The site percolation model is constructed by
occupying sites on a lattice with a given common probability $p$.
Let an occupied site be denoted 'plus' and an 'empty' site denoted
'minus'. For $p$ close to zero the sites are mainly unoccupied and
the lattice is 'minus'. For $p$ close to one the sites are
predominantly occupied and the lattice is 'plus'. At a critical
concentration $p_c$, the percolation threshold, an infinite
cluster of 'plus' sites embedded in the 'minus' background extends
across the lattice. In the scaling limit of vanishing lattice
distance the critical cluster has a fractal boundary which can be
accessed by SLE. Whereas the scaling properties of critical
percolation clusters define a universality class and is
independent of the lattice structure, the critical concentration
$p_c$ in general depends on the lattice. For site percolation on a
triangular lattice the percolation threshold is known to be
$p_c=1/2$.

\begin{figure}
\begin{center}
\includegraphics[width=0.7\hsize]
{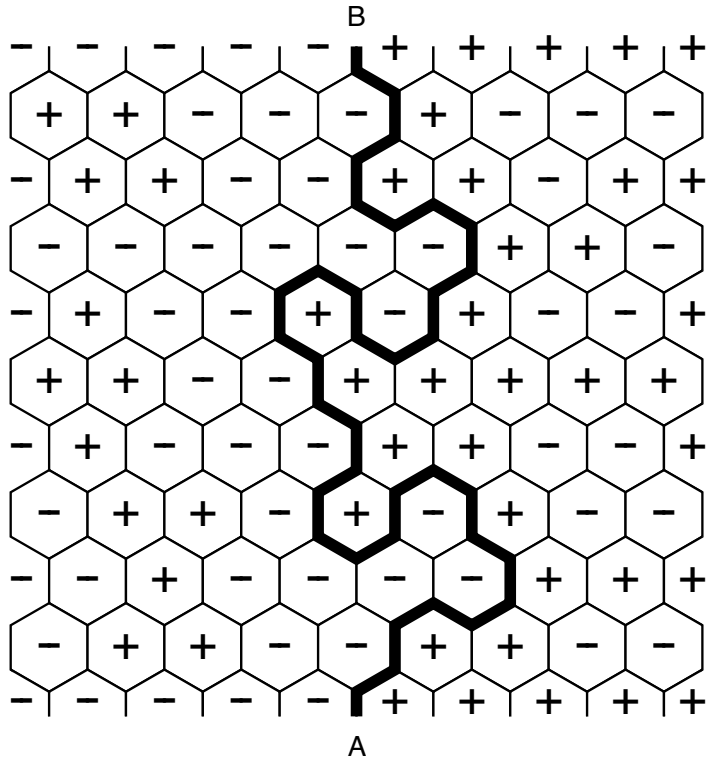}\caption {We depict site
percolation on a triangular lattice in the upped half plane at the
percolation threshold. The critical concentration is $p_c=1/2$.
The occupied sites are denoted 'plus', the empty sites 'minus'.
The boundary conditions enforce a meandering domain wall from  A
to B.} \label{fig1}
\end{center}
\end{figure}

In Fig.\ref{fig1} we depict a realization of site percolation on a
triangular lattice in the upper half plane at the percolation
threshold $p_c=1/2$. The occupied sites are denoted 'plus', the
empty sites 'minus'. By imposing appropriate boundary conditions
we induce a meandering domain wall across the system from A to B.
In the scaling limit the domain wall becomes a fractal
non-crossing critical curve.

\subsection{Ising Model}

The Ising model is probably the simplest interacting many particle
system in statistical physics [8,14,32]. The model has its origin
in magnetism but has become of paradigmatic importance in the
context of phase transitions. The model is defined on a lattice
where each lattice site $i$ is occupied by a single degree of
freedom, a spin variable $\sigma_i$, assuming two values,
$\sigma_i=\pm 1$, i.e., spin up or spin down. In the ferromagnetic
case considered here the spins interact via a short range exchange
interaction $J$ favoring parallel spin alignment. The model is
described by the Ising Hamiltonian
\begin{eqnarray}
H=-J\sum_{\langle ij\rangle}\sigma_i\sigma_j, \label{ham}
\end{eqnarray}
where $\langle ij\rangle$ indicates nearest neighbor spin sites $i$
and $j$. The statistical weight or probability of a specific spin
configuration $\{\sigma_i\}$ is determined by the Boltzmann factor
\begin{eqnarray}
P(\{\sigma_i\})=\exp[-H/kT]/Z, \label{boltz}
\end{eqnarray}
where $T$ is the temperature and $k$ Boltzmann's constant. The
partition function $Z$ has the form
\begin{eqnarray}
Z=\sum_{\{\sigma_i\}}\exp(-H/kT), \label{part}
\end{eqnarray}
yielding the thermodynamic free energy $F$ according to $F=-kT\log
Z$. The entropy is given by $S=-dF/dT$ and the energy follows from
$F=E-TS$. The magnetization or order parameter and correlations
are given by
\begin{eqnarray}
m=\sum_{\{\sigma_i\}}\sigma_i P(\{\sigma_i\})~~\text{and}~~
\langle\sigma_i\sigma_j\rangle=\sum_{\{\sigma_i\}}\sigma_i\sigma_j
P(\{\sigma_i\}), \label{order-corr}
\end{eqnarray}
respectively.

The Ising model possesses a phase transition at a critical
temperature $T_c$ from a disordered paramagnetic phase above $T_c$
with vanishing order parameter $m=0$ to a ferromagnetic ordered
phase below $T_c$ with non-vanishing order parameter $m\neq 0$. At
the critical temperature $T_c$ the order parameter vanishes like
$m\sim|T-T_c|^\beta$ with critical exponent $\beta$. The correlation
function $\langle\sigma_i\sigma_j\rangle$ monitoring the spatial
organization of ordered domains behaves like
\begin{eqnarray}
\langle\sigma_i\sigma_j\rangle\sim\frac{\exp[-|i-j|/\xi]}{|i-j|^\eta},
\label{corr2}
\end{eqnarray}
where $\eta$ is a critical exponent and the correlation length
$\xi$ scales like $\xi\sim|T-T_c|^{-\nu}$ with critical exponent
$\nu$. At the critical point the correlation length $\xi$ diverges
and the Ising model becomes scale invariant with an algebraically
decaying correlation function
$\langle\sigma_i\sigma_j\rangle\sim|i-j|^{-\eta}$. Assigning
'plus' to spin up and 'minus' to spin down, Fig.\ref{fig1} also
illustrates a typical configuration of the 2D Ising model on a
triangular lattice at the critical temperature, including an Ising
domain wall from A to B.

\subsection{Critical curves - Exploration}

In the percolation case at the percolation threshold a critical
curve is induced by fixing the boundary conditions. Occupying
sites from $A$ to $B$ along the right hand side of the boundary
and assigning empty sites along the left hand side of the boundary
from $A$ to $B$ a critical curve will meander across the system
from $A$ to $B$. Imagining painting the two sides of the curve,
one side is painted with occupied sites, the other side with empty
sites. Typically the curve meanders on all scales but by
construction does not cross itself. For later purposed the
configuration is depicted in Fig. \ref{fig1}.

\begin{figure}
\begin{center}
\includegraphics[width=.7\hsize]
{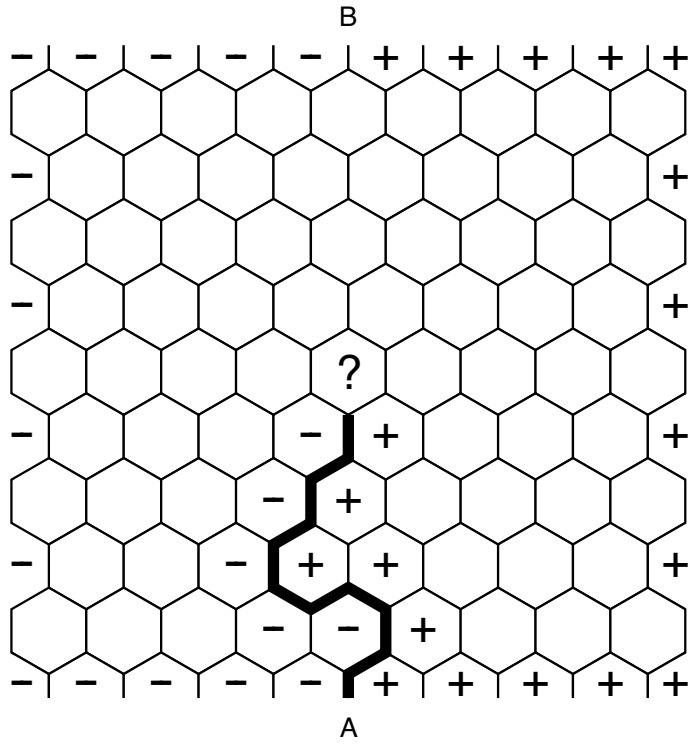}\caption {We depict the
growth process in the percolation case. The percolation threshold
is at $p_c=1/2$. The interface imposed by the boundary conditions
originates at the boundary point $A$ and progresses towards the
boundary point $B$.} \label{fig2}
\end{center}
\end{figure}

In order to make contact with SLE we observe that a critical
interface in the percolation case also can be constructed by an
exploration process. We initiate the curve at the boundary point $A$
and toss a coin. In the case of 'head' the site or hexagon in front
is chosen to be occupied and the path bends left; in the event of
'tail' the hexagon in front is left unoccupied and the path bends
right. In this manner a critical meandering non-crossing curve is
generated terminating eventually at $B$. The percolation growth
process is depicted in Fig. \ref{fig2}, where we for clarity only
have indicated the sites involved in the growth process. We note
that since there is no interaction between the sites the path
depends entirely on the local properties.

In the case of the Ising model a critical interface or domain
walls at the critical point is again fixed by assigning
appropriate boundary conditions with spin up along the boundary
from $A$ to $B$ and spin down from $B$ to $A$. The critical curve
can be constructed in two ways: Globally or by an exploration
process. In the global case we generate a spin configuration by
means of a Monte Carlo simulation, i.e., perform a biased
importance sampling implementing the probability distribution in
Eq. (\ref{boltz}), and identify a critical interface.
Alternatively, we can generate an interface by an exploration
process like in the percolation case, occupying a site $i$
according to the weight $(1/2)(1+\langle\sigma_i\rangle)$, where
$\langle\sigma_i\rangle$ is evaluated in the domain with the spins
along the interface fixed, see again Fig. \ref{fig2}.

\subsection{Distributions - Markov properties - Measures}

In the case of random walk the number of walks of length $L$ grows
like $\mu^L$, where $\mu$ is a lattice dependent number, i.e., at
each step there are $\mu$ lattice-dependent choices for choosing a
direction of the next step. Consequently, the weight or
probability of a particular walk of length $L$ is proportional to
$\mu^{-L}$,
\begin{eqnarray}
P(L)\sim \mu^{-L}, \label{prob}
\end{eqnarray}
and all walks of a given length have the same weight. We note here
the important Markov property, characteristic of random walk,
which can be formulated in the following manner. Assume that the
first part $\gamma'$ of the walk has taken place and thus
conditions the remaining part $\gamma$ of the walk. In a
suggestive notation the conditional distribution is given by
$P(\gamma|\gamma') = P(\gamma\gamma')/P(\gamma')$. The Markov
property implies that the conditional probability is  equal to the
probability of the walk $\gamma$ in a new domain where the first
part of the walk $\gamma'$ has been removed, i.e., the identity
\begin{eqnarray}
P(\gamma|\gamma')=P'(\gamma),~~~~ \text{(Markov property)}
\label{mar}
\end{eqnarray}
where the prime refers to the new domain. The Markov property is
self-evident for random walk and follows directly from Eq.
(\ref{prob}), i.e., $P(\gamma|\gamma') = P(\gamma\gamma')/P(\gamma')
= \mu^{-(L+L')}/\mu^{-L'} = \mu^{-L}=P'(\gamma),$ where $L$ and $L'$
are the lengths of segments $\gamma'$ and $\gamma$, respectively.

In the scaling limit the lattice distance goes to zero whereas the
length of the walk diverges. Consequently, the distribution
diverges and must be replaced by an appropriate probability
measure [3,6,21]. However, in order to define the probability
distribution or measure in the scaling limit we shall assume that
the Markov property continues to hold and interpret the $P$ in Eq.
(\ref{mar}) as probability measures. The Markov property is
essential in carrying over the lattice probability distributions
in the scaling limit.

For the critical interfaces defined by an exploration processes in
the case of site percolation, the Markov property follows by
inspection since the propagation of the interface is entirely
determined by the local process of occupying the next site with
probability $1/2$. This locality property is specific for
percolation which has a geometrical phase transition and does in the
SLE context, to be discussed later, determine the scaling
universality class.

In the case of an interface in the Ising model the Markov property
also holds, but since the spins interact a little calculation is
required [6]. Consider an interface $\gamma$ defined by an
exploration process. According to the rules of statistical mechanics
the probability distribution for the interface $\gamma$ is given by
\begin{eqnarray}
P(\gamma)=\frac{Z(\gamma)}{Z}. \label{probis}
\end{eqnarray}
Here $Z$ is the full partition function defined in Eq. (\ref{part})
with appropriate boundary conditions imposed. $Z(\gamma)$ is the
partial partition function with the spins associated with the
interface $\gamma$ fixed,
\begin{eqnarray}
Z(\gamma)=\sum_{\{\sigma_i\},\gamma}\exp[-H(\gamma)/kT].
\label{part2}
\end{eqnarray}
The Hamiltonian $H(\gamma)$ inferred from Eq. (\ref{ham}) is the
energy of a spin configuration with the spins determining $\gamma$
fixed; $\{\sigma_i\},\gamma$ indicate the configurations to be
summed over. Evidently, we have the identity $Z=\sum_\gamma
Z(\gamma)$, i.e., $\sum_\gamma P(\gamma)=1$.

Whereas in the random walk case we only considered an individual
path and in the percolation case the interface only feels the nearby
sites, an Ising interface is imbedded in the interacting spin
systems and we have to define the Markov property more precisely
with respect to a domain $D$. Consider an interface across the
domain $D$ from $A$ to $B$ and assume that the last part $\gamma$ is
conditioned on the determination of the first part $\gamma'$, i.e.,
given by the distribution $P_D(\gamma|\gamma')$. Next imagine that
we cut the domain $D$ along the interface $\gamma'$, i.e., break the
interaction bonds between the spins determining $\gamma'$. The right
and left face of $\gamma'$ can then be considered part of the domain
boundary and the Markov property states that the distribution of
$\gamma$ in the cut domain $D\setminus\gamma'$ (i.e., $D$ minus
$\gamma'$)  equals $P_D(\gamma|\gamma')$,
\begin{eqnarray}
P_D(\gamma|\gamma')=P_{D\setminus\gamma'}(\gamma).~~~~\mbox{(Markov
property)} \label{mar2}
\end{eqnarray}
In order to demonstrate Eq. (\ref{mar2}) we use the definition in
Eq. (\ref{probis}). The conditional probability
$P_D(\gamma|\gamma')=P_D(\gamma\gamma')/P_D(\gamma')=
(Z_D(\gamma\gamma')/Z_D)/(Z_D(\gamma')/Z_D) =
Z_D(\gamma\gamma')/Z_D(\gamma')$. Correspondingly, the
distribution in the cut domain $D\setminus\gamma'$ is
$P_{D\setminus\gamma'}(\gamma) =
Z_{D\setminus\gamma'}(\gamma)/Z_{D\setminus\gamma'}$. However, it
follows from the structure of the partition function in Eqs.
(\ref{ham} - \ref{part}) that $Z_{D\setminus\gamma'}=
\exp[E(\gamma')/kT]Z_D(\gamma')$ and
$Z_{D\setminus\gamma'}(\gamma)=
\exp[E(\gamma')/kT]Z_D(\gamma\gamma')$, where $E(\gamma')$ is the
energy of the broken bonds. By insertion the interface Boltzmann
factors cancel out and we obtain Eq. (\ref{mar2}) expressing the
Markov property.

\section{Conformal Invariance}

In the complex plane analysis implies geometry. The representation
of a complex number $z=x+iy$ directly associates complex function
theory with 2D geometrical shapes. This connection is of importance
in mathematical physics in for example 2D electrostatics and 2D
hydrodynamcs . In the context of SLE Riemann's mapping theorem plays
an essential role [1].

\subsection{Conformal Maps}

\begin{figure}
\begin{center}
\includegraphics[width=0.7\hsize]
{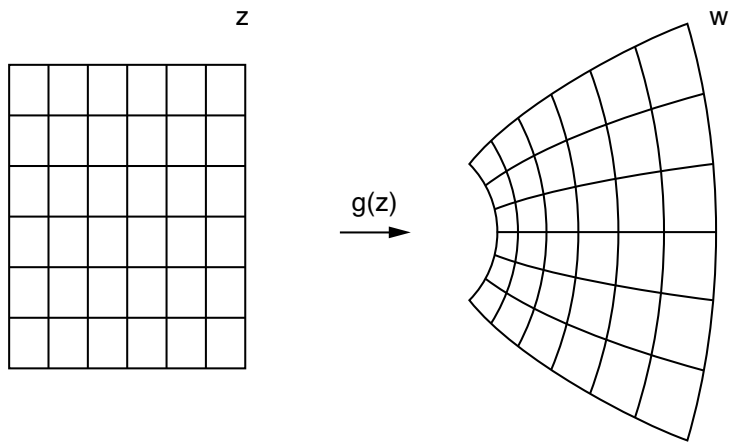}\caption {We depict a
conformal transformation from the complex $z$ plane to the complex
$w$ plane. We note the angle-preserving property, i.e., a
shear-free transformation. The map in the figure is given by
$w=z^2$.} \label{fig3}
\end{center}
\end{figure}

Briefly, Riemann's mapping theorem [2] states that any simply
connected domain, i.e., topologically deformable to a disk, in the
complex plane $z$ can be uniquely mapped to a unit disk $|w|<1$ in
the complex $w$ plane by mean of a complex function $w=g(z)$. By
combining complex functions we can map any simply connected domain
to any other simply connected  domain. For example, if $g_1(z)$
and $g_2(z)$ map domains $D_1$ and $D_2$ to the unit disk,
respectively, then $g_2^{-1}(g_1(z))$ maps $D_1$ to $D_2$; here
$g_2^{-1}$ is the inverse function of $g_2, i.e.,
g_2^{-1}(g_2(z))=z$. As an example, the transformation
$g(z)=i(1+z)/(1-z)$ maps the the unit disk centered at the origin
to the upper half plane. Likewise, the M\"obius transformation
$g(z)=(az+b)/(cz+d)$ determined by four real parameters,
$ad-bc>0$, maps the upper half plane onto itself. Conformal
transformations are angle-preserving and basically correspond to a
combination of a local rotation, local translation, and local
dilatation. In terms of an elastic medium picture conformal
transformations are equivalent to deformations without shear
(Landau 1959). Expressing $g(z)$ in terms of its real and
imaginary parts, $g(z)=u(x,y)+iv(x,y)$, the Cauchy-Riemann
equations $\partial u/\partial x= \partial v/\partial y$ and
$\partial u/\partial y= -\partial v/\partial x$ hold implying that
$u$ and $v$ are harmonic functions satisfying Laplace's equations
$\nabla^2 u=0$ and $\nabla^2 v=0$, $\nabla^2=\partial^2/\partial
x^2+\partial^2/\partial y^2$. In Fig.~\ref{fig3} we have depicted
a conformal angle-preserving transformation effectuated by the
complex function $g(z)$ from the complex $z$ plane to the complex
$w$ plane.

\subsection{Measures - Conformal Invariance}

Whereas the Markov property discussed above holds for lattice
curves even away from criticality, we here want to assume another
property which only holds in the scaling limit at the critical
point, namely conformal invariance. In the scaling limit we
anticipate that the probability measure $P(\gamma)$ for an
interface $\gamma$ is invariant under a conformal transformation.
More precisely, consider a lattice model, say the Ising or
percolation model, and specify two domains $D$ and $D'$ on the
lattice. Next, consider an interface, cluster boundary or domain
wall $\gamma$ from the boundary points $A$ and $B$ across the
domain $D$. In terms of the partition functions the probability
distribution for $\gamma$ is given by Eq. (\ref{probis}). We now
perform the scaling or continuum limit of the lattice model
keeping the domains $D$ and $D'$ fixed. The continuous random
interface approaches its scaling form and is characterized by the
measure $P_D(\gamma)$. At the critical point we assume that the
interface is scale invariant under the larger symmetry of
conformal transformations. The next step is to consider a specific
conformal transformation $g(z)$ which according to Riemann's
theorem precisely maps domain $D$ to domain $D'$, i.e., $D'=g(D)$
and the interface to $g(\gamma)$. The assumption of conformal
invariance then states that the probability measure $P$ is
invariant under this transformation expressing the scale
invariance, i.e.,
\begin{eqnarray}
P_D(\gamma)=P_{g(D)}(g(\gamma)).~~~~\mbox{(Conformal property)}
\label{conf}
\end{eqnarray}
Both the Markov property and the conformal property are sufficient
in combination with Loewner evolution to determine the measures in
the scaling limit.

\section{Loewner Evolution}

The original motivation of Loewner's work was to examine the so
called Bieberbach conjecture which states that $|a_n|\leq n$ for
the coefficients in the Taylor expansion $f(z)=\sum_{n=0}a_nz^z$.
The conjecture was proposed in 1916 and finally proven in 1984 by
de Branges [2,19,57]. For that purpose Loewner [76] considered
growing parametrized conformal maps to a standard domain. In the
present context Loewner's method allow us to access growing shapes
in 2D in an indirect manner by means of a 'time dependent'
conformal transformation $g_t(z)$.

\subsection{Growing Stick}

\begin{figure}
\begin{center}
\includegraphics[width=0.7\hsize]
{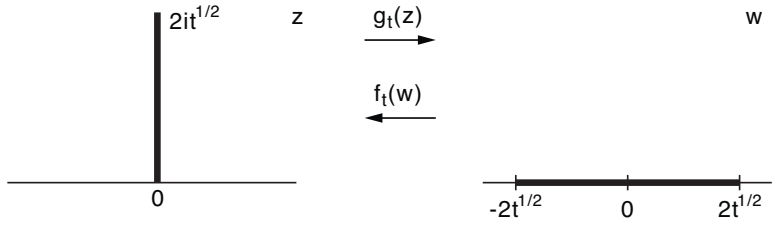}\caption {We depict the
growing stick corresponding to the conformal transformation
$w=\sqrt{z^2+4t}$. The vertical cut in the $z$ plane extends from
the origin to the point $(0,2it^{1/2})$. The right and left faces
of the cut are mapped to the real axis from $-2t^{1/2}$ to
$+2t^{1/2}$, the endpoint to the origin, in the complex $w$
plane.} \label{fig4}
\end{center}
\end{figure}

Before we address the derivation of the Loewner equation let us
consider the specific conformal transformation
\begin{eqnarray}
g_t(z)=\sqrt{z^2+4t}. \label{stick}
\end{eqnarray}
For $t=0$ we have $g_0(z)=z$, i.e., the identity map. Likewise, for
$z\rightarrow\infty$ we obtain
\begin{eqnarray}
g_t(z)\sim z+\frac{2t}{z}, \label{asym}
\end{eqnarray}
showing that far away in the complex plane we again have the
identity map. The coefficient in the next leading term, $C_t/z$,
is called the capacity $C_t$; here parametrized by the 'time
variable', $t=C_t/2$. The map (\ref{stick}) has a branch point at
$z=2it^{1/2}$ and it follows by inspection that $g_t$ maps the
upper half plane minus a stick from the origin $\bf 0$ to
$2it^{1/2}$ back to the upper half plane. From the inverse map
$f_t(w)=g_t^{-1}(w)$,
\begin{eqnarray}
f_t(w)=\sqrt{w^2-4t}, \label{inv}
\end{eqnarray}
we infer that the right face of the stick is mapped to the real
axis from $0$ to $2t^{1/2}$, the tip to the origin, and the left
face to the interval $-2t^{1/2}$ to $0$. Under the map the growing
stick thus becomes part of the boundary in the $w$ plane. The
growing stick is depicted in Fig.~\ref{fig4}.

The stick shows up as an imaginary contribution along the interval
$-2t^{1/2}$ to $2t^{-1/2}$. More precisely, since $f_t(w)$ is
analytic in the upper half plane, implementing the asymptotic
behavior $f_t(w)\sim w$ for $w\rightarrow\infty$, and using
Cauchy's theorem, we obtain the dispersion relation or spectral
representation
\begin{eqnarray}
f_t(w)=w-\int\frac{d\omega}{\pi}\frac{A_t(\omega)}{w-\omega},
\label{spec}
\end{eqnarray}
with spectral weight $A_t(\omega)$. In the case of the growing stick
we find $A_t(\omega)=(4t-\omega^2)^{1/2}$ for $\omega^2<4t$ and
otherwise $A_t(\omega)=0$. Using $1/(\omega+i\epsilon)= \text{P}~
1/\omega-i\pi\delta(\omega)$ (P denotes principal value) we also
have $\text{Im} f_t(\omega)=A_t(\omega)$ and $\text{Re} f_t(\omega)
= \omega - \text{P}\int(d\omega'/\pi)A_t(\omega')/(\omega-\omega')$.
The time dependent spectral weight $A_t(\omega)$ thus characterizes
the growing stick. With the chosen parametrization we also have the
sum rule
\begin{eqnarray}
\int\frac{d\omega}{\pi}A_t(\omega)=C_t=2t. \label{sum}
\end{eqnarray}
Finally, we note that the map $g_t$ satisfies the equation of motion
\begin{eqnarray}
\frac{dg_t(z)}{dt}=\frac{2}{g_t(z)}, \label{eqmo}
\end{eqnarray}
i.e., solving Eq. (\ref{eqmo}) with the initial condition
$g_0(z)=z$ and the boundary condition $g_t(z)\sim z$ for
$z\rightarrow\infty$ we arrive at the map in Eq. (\ref{stick}).

\subsection{Loewner Equation}

The growing stick nicely illustrates the idea of accessing a growing
shape indirectly by the application of Riemann's theorem mapping the
domain adjacent to the shape to a standard reference domain, here
the upper half plane. This so-called uniformizing map effectively
absorbs the shape and encodes the information about the shape into
the spectral weight $A_t(\omega)$ along the real axis.

\begin{figure}
\begin{center}
\includegraphics[width=0.7\hsize]
{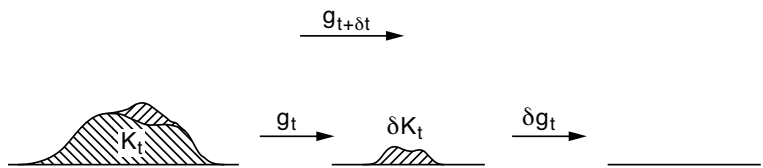}\caption{The combination of
maps involved in the derivation of the Loewner equation. First the
map $g_t$ eliminates the hull $K_t$. Subject to the growth in the
time interval $\delta t$ the incremental hull $\delta K_t$ is
subsequently absorbed by the infinitesimal map $\delta g_t$.
Correspondingly, the hull $K_{t+\delta t}$ is absorbed by the map
$g_{t+\delta t}$ in one step. } \label{fig5}
\end{center}
\end{figure}

Let us consider a general shape or hull $\bf K_t$ in the upper
half plane $\bf H$. Together with the real axis the shape form
part of the boundary of the domain $\bf D$. In other words, the
domain in question is the upper half plane $\bf H$ with the shape
$\bf K_t$ subtracted, $\bf D = \bf H\setminus\bf K_t$. Applying
Riemann's theorem we map the simply connected domain $\bf D$ back
to the upper half plane $\bf H$ by means of the conformal
transformation $g_t(z)$, i.e., $g_t$  absorbs the shape $\bf K_t$.
Imagine that the shape grows a little bit further from $\bf K_t$
to $\bf K_{t+\delta t}=\bf K_t+\delta\bf K_t$, where $\bf K_t$ is
now part of $\bf K_{t+\delta t}$; $\delta\bf K_t$ is the shape
increment. Correspondingly, the map $g_{t+\delta t}$ is designed
to absorb $\bf K_{t+\delta t}$, i.e., $\bf H\setminus\bf
K_{t+\delta t}\rightarrow \bf H$ by means of the map $g_{t+\delta
t}$. We now carry out the elimination in two ways. Either we
absorb $\bf K_t$ by means of the map $g_t$ and subsequently
$\delta\bf K_t$ by means of the map $\delta g_t$ or we absorb $\bf
K_{t+\delta t}$ directly in one step by means of the map
$g_{t+\delta t}$. Consequently, combining maps we have
$g_{t+\delta t}(z)=\delta g_t(g_t(z))$ or $g_t(z)=\delta
g_t^{-1}(g_{t+\delta t}(z))$. Since $\delta g_t^{-1}(w)$ is
analytic in $\bf H$ we obtain the spectral representation
\begin{eqnarray}
\delta g_t^{-1}(w)=w-\int\frac{d\omega}{\pi}\frac{\delta
A_t(\omega)}{w-\omega}, \label{spec2}
\end{eqnarray}
with infinitesimal spectral weight $\delta A_t$, or inserting
$w=g_{t+\delta}(z)$
\begin{eqnarray}
g_t(z)=g_{t+\delta t}(z)-\int\frac{d\omega}{\pi}\frac{\delta
A_t(\omega)}{g_{t+\delta t}-\omega}. \label{inter}
\end{eqnarray}
The last step is to set $\delta A_t(\omega)=\rho_t(\omega)\delta t$,
yielding a differential equation for the evolution of the map $g_t$
eliminating the shape $\bf K_t$,

\begin{eqnarray}
\frac{dg_t(z)}{dt}=
\int\frac{d\omega}{\pi}\frac{\rho_t(\omega)}{g_t(z)-\omega}.
\label{lchain}
\end{eqnarray}
Specifying the weight or measure $\rho_t(\omega)$ along the real
$\omega$-axis this equation determines, through the uniformizing
map $g_t$, how the shape $\bf K_t$ grows. The spectral weight
encodes the 2D shape into the real function $\rho_t(\omega)$. Note
that since $\rho_t(\omega)$ is not specified and can depend
nonlinearly on the map, Eq. (\ref{lchain}) still represent a
highly nonlinear problem. Invoking the asymptotic condition
$g_t(z)\sim z+C_t/z$, where $C_t$ is the time dependent capacity
we infer $dC_t/dt=\int(d\omega/\pi)\rho_t(\omega)$ or since
$\rho_t(\omega)=dA_t(\omega)/dt$ the sum rule in Eq. (\ref{sum}).
The conformal mapping procedure is depicted in Fig.~\ref{fig5}.

\begin{figure}
\begin{center}
\includegraphics[width=0.7\hsize]
{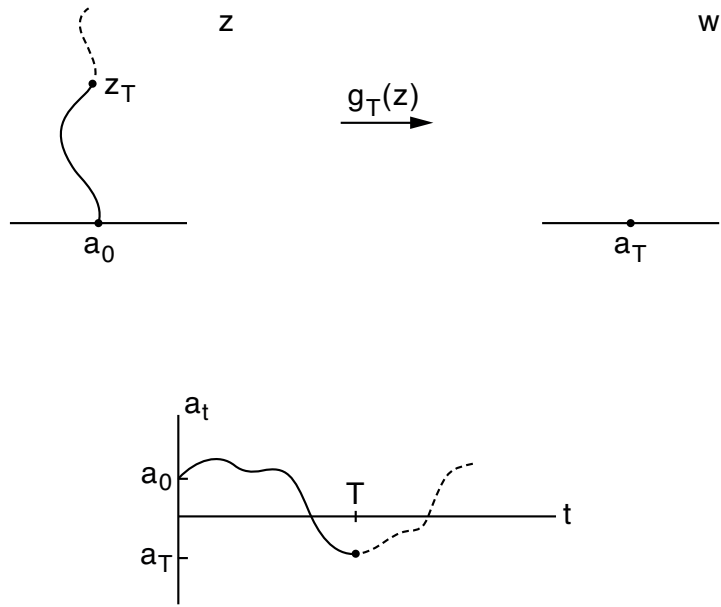}\caption {The mechanism in
the Loewner equation. The curve in the upper half complex $z$
plane generated by the Loewner equation is mapped onto a finite
but growing segment of the real axis of the complex $w$ plane. The
endpoint $z_T$ is mapped to the real number $a_T$. As $a_t$
develops in time and makes excursions along the real axis the
endpoint $z_t$ of the curve grows into the upper half plane.}
\label{fig6}
\end{center}
\end{figure}

In the special case where the growth takes place at a point the
equation simplifies considerably. Assuming that the spectral
weight is concentrated at the point $\omega=a_t$, where $a_t$ is a
real function of $t$ and setting
$\rho_t(\omega)=2\pi\delta(\omega-a_t)$ we arrive at the Loewner
equation
\begin{eqnarray}
\frac{dg_t(z)}{dt}= \frac{2}{g_t-a_t}. \label{loeweq}
\end{eqnarray}
The Loewner equation describes the growth of a curve or trace
$\gamma_t$ with endpoint $z_t$, $0<t<\infty$, in the upper half
complex $z$-plane. The time-dependent conformal transformation
$g_t$ maps the simply connected domain $\bf H\setminus\gamma_t$,
i.e., the half plane excluding the curve $\gamma_t$ back to the
$w$ half plane. At a given time instant $t$ the tip of the curve
$z_t$ is determined by $g_t(z_t)=a_t$, i.e., the point where Eq.
(\ref{loeweq}) develops a singularity. The topological properties
and shape of the curve are encoded in the real function $a_t$
which lives on the real axis in the $w$-plane. As $a_t$ develops
in time the tip of the curve $z_t$ determined by
$z_t=g_t^{-1}(a_t)$ traces out a curve. Since the domain $\bf
H\setminus\gamma_t$ must be simply connected for the Riemann
theorem to apply the curve or trace cannot cross itself or cross
the real axis. Whenever the curve touches or intersect itself or
the real axis the enclosed part will be excluded from the domain.
In other, words, during the time progression the curve effectively
absorbs part of the upper half plane. It is a deep property of
Loewner evolution that the topological properties of a 2D
non-crossing curve are entirely encoded by the real function
$a_t$. The encoding works both ways: A given 2D non-crossing curve
$\gamma_t$ corresponds to a specific real function $a_t$, a given
real function $a_t$ yields a specific 2D non-crossing curve
$\gamma_t$. A continuous $a_t$ will yield a continuous curve
$\gamma_t$. A discontinuous $a_t$ in general gives rise to
branching. Whether or not the curve intersects or touches itself
is determined by the singularity structure of the drive $a_t$. In
the case where the H\"older condition $\lim_{\tau\rightarrow
0}|(a_{t+\tau}-a_\tau)/\tau^{1/2}|$ is greater that $4$ we have
self-intersection. Note again that since the curve is defined
indirectly by the singularity structure in Eq. (\ref{loeweq}) we
cannot easily identify a curve parametrization and for example
determine a tangent vector, etc. The mechanism underlying the
Loewner equation is shown in Fig.~\ref{fig6}.

\subsection{Exact Solutions}

In a series of simple cases one can solve the Loewner equation
analytically [59,60]. For vanishing drive $a_t=0$ we obtain the
growing vertical stick discussed above. Correspondingly, a constant
drive $a_t=a$ yields a vertical stick growing up from the point $a$
on the real axis.

In the case of a linear drive $a_t=t$ the tip of the curve $z_t$
is given by $z_t=2-2\phi_t\cos\phi_t+2i\phi_t$, where the phase
$\phi_t$ is determined from the equations: $2\ln
r_t-r_t\cos\phi_t=2\ln 2+t-2$ and $r_t=2\phi_t/\sin\phi_t$. By
inspection $\phi_0=0$ and $\phi_\infty=\pi$. The curve thus
approaches the asymptote $2\pi i$ for $t\rightarrow\infty$. For
small $t$ analysis yields $z_t\sim (2/3)t+2i\sqrt t$, i.e., the
trace approaches the origin with infinite slope. The square root
drives $a_t=2\sqrt{\kappa t}$ and $a_t=2\sqrt{\kappa(1-t)},~~
0<t<1$ with a finite-time singularity can also be treated. In the
first case, $a_t = 2\sqrt{\kappa t}$, the trace is a straight line
$z_t= B\exp(i\phi)\sqrt t$ forming the angle $\phi$ with respect
to the real axis. The amplitude $B$ and phase $\phi$ depend on the
parameter $\kappa$. The angle $\phi =
(\pi/2)(1-\kappa^{1/2}/(\kappa+4)^{1/2})$. For $\kappa=0$,
$\phi=\pi/2$ and we recover the perpendicular stick; for
$\kappa\rightarrow\infty$, $\phi\rightarrow 0$ and the angle of
intersection decreases to zero. In the second case, $a_t =
2\sqrt{\kappa(1-t)}$, the behavior of the trace is more complex.
For $0<\kappa<4$ the trace forms a finite spiral in the upper half
plane; for $\kappa=4$ the trace has a glancing intersection with
the real axis. For $4<\kappa<\infty$ the trace hits the real axis
in accordance with H\"older condition discussed above.

\section{Stochastic Loewner Evolution}

After these preliminaries we are in position to address stochastic
Loewner evolution (SLE). The essential observation made by Oded
Schramm [82] within the context of loop erased random walk was that
the Markov and conformal properties of the measures or probability
distributions for random curves generated by Loewner evolution imply
that the random drive $a_t$ must be proportional to an unbiased 1D
Brownian motion.

\subsection{Schramm's Theorem}
The Loewner equation (\ref{loeweq}) generates a non-crossing curve
in the upper half plane $\bf H$ originating at the origin $\bf O$,
given a continuous function $a_t$ with initial value $a_0=0$. As
$a_t$ develops in time the tip of the curve $z_t$ determined by
the condition $g_t(z_t)=a_t$ traces out a curve or trace. In the
case where $a_t$ is a continuous random function the Loewner
equation (\ref{loeweq}) likewise becomes a stochastic equation of
motion yielding a stochastic map $g_t(z)$. As a result the trace
determined by $g_t(z_t)=a_t$ or $z_t=g_t^{-1}(a_t)$ is a random
curve. The issue is to establish a contact between the exploration
processes defining interfaces in the lattice models, the scaling
limit of these curves, and the curves generated by SLE. In the
scaling limit we thus invoke the two properties discussed above:
i) the Markov property in Eq. (\ref{mar}) and ii) the conformal
property in Eq. (\ref{conf}).

In order to demonstrate the surprising property that the Markov
and conformal properties in combination imply that $a_t$ must be a
1D Brownian motion we focus on chordal SLE which applies to a
random curve or trace connecting two boundary points. Since the
probability distribution or measure $P(\gamma)$ on the random
curve $\gamma$ using property ii) is assumed to be conformally
invariant and since we by Riemann's theorem can map any simply
connected domain to the upper half plane by mean of a conformal
transformation, we are free to consider curves in the upper half
plane from the origin $\bf O$ to $\infty$ parametrized with a time
coordinate $0<t<\infty$.

\begin{figure}
\begin{center}
\includegraphics[width=0.7\hsize]
{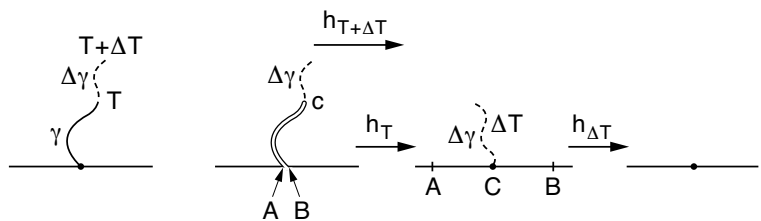} \caption {The figure
depicts the construction in the derivation of SLE. The first step
implements the Markov property by turning the curve $\gamma$ into
a cut. Subsequently, the conformal transformation $h_T$ maps
$\gamma$ back to the origin. Finally, the map $h_{\Delta T}$ maps
the segment $\Delta\gamma$ to the origin. The complete process is
also implemented by $h_{T+\Delta T}$. The combination of the
Markov property and conformal invariance implies that $a_t$
performs a Brownian motion} \label{fig7}
\end{center}
\end{figure}

Imagine that we grow the curve from time $t=0$ to time $T$ driven
by the function $a_t$, $0<t<T$. The curve or trace is generated by
the Loewner equation (\ref{loeweq}) with boundary condition
$a_0=0$ and the trace $z_t$ by $g_t(z_t)=a_t$ or
$z_t=g_t^{-1}(a_t)$. With the chosen time parametrization we have
$g_t(z)\sim z+2t/z$ for $z\rightarrow\infty$ in the upper half
plane. The map $g_t$ thus uniformizes the trace, i.e., the tip
$z_t$ is mapped to $a_t$ on the real axis in the $w$-plane. In
order to invoke the Markov property we let the curve grow the time
increment $\Delta T$ corresponding to the curve segment
$\Delta\gamma$. The Markov property then implies that the
distribution on $\Delta\gamma$ conditioned on the distribution on
$\gamma$ is the same as the distribution on  $\Delta\gamma$ in the
cut domain ${\bf H}\setminus\gamma$, i.e., the domain with the
curve $\gamma$ deleted; this stage is illustrated in
Fig.~\ref{fig7}.

Next, in order to implement the conformal property we shift the
image by $a_T$ in such a way that the curve segment $\Delta\gamma$
again starts at the origin $\bf O$. This is achieved by using the
map $h_T=g_T-a_T$ which since $g_T(z_T)=a_T$ maps the tip $z_T$
back to the origin; this construction is also depicted in
Fig.~\ref{fig7}. Moreover, the asymptotic behavior of $h_T$ for
large $z$ is $h_T(z)\sim z-a_T-2T/z$. Since the measure by
assumption is unchanged under the conformal transformation $h_T$
we infer that $\Delta\gamma$ growing the time $\Delta T$ from the
origin has the same distribution as the segment $\Delta\gamma$
grown from time $T$ to time $T+\Delta T$ conditioned on the
segment $\gamma$ grown up to time $T$. Moreover, since the segment
$\gamma$ from $\bf O$ $(AB)$ to $C$ subject to the Markov property
has become part of the boundary, as shown in Fig.~\ref{fig7}, we
also infer that the measure on $\Delta\gamma$ is independent of
the measure on $\gamma$. Finally, applying $h_{\Delta T}$ we map
the segment $\Delta\gamma$ to the origin as a common reference
point as indicated in Fig.~\ref{fig7}.

Since the random curves are determined by the random maps $g_t$
and $h_t$ driven by the random function $a_t$ the issue is how to
transfer the properties of the measure on the curve determined by
the Markov and conformal properties to the measure on the random
driving function $a_t$.

In order to combine the Markov properties arising from the
analysis of the lattice models and the conformal invariance
pertaining to critical random curves, we carry out the following
steps. First we grow the curve $\gamma$ from the origin $\bf O$ to
the tip $z_T$. Implying conformal invariance the curve is then
uniformized back to the origin by means of $h_T$. The next step is
to grow the curve segment $\Delta\gamma$ in time $\Delta T$. This
segment is subsequently absorbed by means of the map $h_{\Delta
T}$. According to the Markov property the distribution of
$\Delta\gamma$ from $\bf O$ to $z_{\Delta T}$ is the same as the
distribution of $\Delta\gamma$ grown from $T$ to $T+\Delta T$
conditioned on $\gamma$ grown from $0$ to $T$. Since the curve
$\gamma$ is determined by the map $h_t$ the distribution is
reflected in $h_t$. In particular, the stochastic properties of
the curve is transferred to the random function $a_t$ generating
the curve by the Loewner evolution. The last step is now to
observe that absorbing the segment $\Delta\gamma$ from $\bf O$ to
$z_{\Delta T}$ by means of $h_{\Delta T}$ is the same
transformation as first applying the inverse map $h_T^{-1}$
followed by the map $h_{T+\Delta T}$; in both cases the end result
is the absorption of the initial curve $\gamma+\Delta\gamma$, see
Fig.~\ref{fig7}. As regards the measure or distribution we have
the equivalence $h_{\Delta T}(z)\approx h_{T+\Delta
T}(h_T^{-1}(z))$. Using the asymptotic form $h_t(z)\sim
z-a_t-2t/z$ we obtain $a_{T+\Delta T}-a_T\approx a_{\Delta T}$;
note that $\approx$ indicates identical distributions or measures.

In conclusion, the Markov property in combination with conformal
invariance implies that $a_{T+\Delta T}-a_T$ is distributed like
$a_{\Delta T}$ (stationarity) and that $a_{\Delta T}$ and
$a_{\Delta T'}$ are independently distributed for non-overlapping
time intervals $\Delta T$ and $\Delta T'$ (Markov property).
Referring to Section III, A on Brownian motion as expressed in
Eqs. (\ref{sta}) and (\ref{ind}), i.e., stationarity and
independence, we infer that $a_t$ is proportional to a Brownian
motion of arbitrary strength $\kappa$, i.e., $a_t=\sqrt\kappa
B_t$. Note that the reflection symmetry $x\rightarrow -x$ holding
in the present context rules out a bias or drift in the 1D
Brownian motion [6,13].

This is the basic conclusion reached by Schramm in the context of
loop erased random walk. Driving Loewner evolution by means of 1D
Brownian motion with different diffusion coefficient or strength
$\kappa$ we generate a one-parameter family of conformally
invariant or scale invariant non-crossing random curves in the
plane.
\subsection{SLE Properties}

Stochastic Loewner evolution is determined by the nonlinear
stochastic equation of motion
\begin{eqnarray}
\frac{dg_t}{dt}=\frac{2}{g_t-a_t}, ~~a_t=\sqrt\kappa B_t.
\label{sle}
\end{eqnarray}
In the course of time $a_t$ performs a 1D Brownian motion on the
real axis starting at the origin $a_0=0$. $a_t$ is a random
continuous function of $t$ and distributed according to $\sqrt\kappa
B_t$. More precisely, $a_t$ is given by the Gaussian distribution
\begin{eqnarray}
P(a,t)=(2\pi t)^{-1/2}\exp[-a^2/2\kappa t], \label{gauss}
\end{eqnarray}
with correlations
\begin{eqnarray}
\langle(a_t-a_s)^2\rangle=\kappa|t-s|. \label{corr3}
\end{eqnarray}
First we notice that a constant shift of the drive $a_t$,
$a_t\rightarrow a_t+b$, is readily absorbed by a corresponding
shift of the map, $g_t\rightarrow g_t+b$. Moreover, using the
scaling property of Brownian motion, $B_{\lambda^2t}=\lambda B_t$,
following from e.g. Eq. (\ref{corr}), we have
$a_{\lambda^2t}=\lambda a_t$ and we conclude from Eq. (\ref{sle})
that $g_t(z)$ has the same distribution as
$(1/\lambda)g_{\lambda^2t}(\lambda z)$, i.e., $g_{\gamma^2
t}(\gamma z)\approx \gamma g_t(z)$. Note that this dilation
invariance is consistent since the origin $z=0$ and $z=\infty$,
the endpoints of curves, are preserved. Note also that the
strength of the drive $\kappa$ is an essential parameter which
cannot be scaled away.

\subsection{Curves - Hulls - Bessel Process}

For vanishing drive, $\kappa=0$, the SLE yields a non random
vertical line from $z=0$, i.e., the growing stick discussed in Sec.
V.A. As we increase $\kappa$ the curve becomes random with
excursions to the right and to the left in the upper half plane. Up
to a critical value of $\kappa$ the random curve is simple; i.e.,
non-touching or non self-intersecting. At a critical value of
$\kappa$ the Brownian drive is so strong that the curve begins to
intersect itself and the real axis. These intersection take place on
all scales since the curve is self-similar or scale invariant.
Denoting the curve by $\gamma_t$ we observe that since Riemann's
theorem uniformizing $\bf H\setminus\gamma_t$ to $\bf H$ only
applies to a simply connected domain, the regions enclosed by the
self-intersections do not become uniformized but are effectively
removed from $\bf H$. The curve $\gamma_t$ together with the
enclosed parts is called the hull $\bf K_t$ and the mapping theorem
applies to $\bf H\setminus\bf K_t$.

In order to analyze the critical value of $\kappa$ we consider the
stochastic equation for $h_t(z)=g_t(z)-a_t$. From Eq. (\ref{sle}) it
follows that
\begin{eqnarray}
\frac{dh_t}{dt}=\frac{2}{h_t}+\xi_t, \label{sle2}
\end{eqnarray}
where $\xi_t=-da_t/dt$ is white noise with correlations
$\langle\xi_t\xi_s\rangle = \kappa\delta(t-s)$.

The nonlinear complex Langevin equation (\ref{sle2}) maps the tip
of the curve $z_t$ back to the origin. Likewise $\bf
H\setminus\gamma_t$ is mapped onto $\bf H$. A point $x$ on the
real axis is mapped to $x_t=h_t(x)$ where $x_t$ satisfies
\begin{eqnarray}
\frac{dx_t}{dt}=\frac{2}{x_t}+\xi_t. \label{sle3}
\end{eqnarray}
The Langevin equation (\ref{sle3}) is known as the Bessel equation
and governs the radial distance $R$ from the origin of a Brownian
particle in $d$ dimensions.

Introducing $R=(\sum_{i=1}^d B_i^2)^{1/2}$ where $B_i$,
$i=1,\cdots d$, is a 1D Brownian motion with distribution
$P(B,t)=(2\pi\kappa t)^{-1/2}\exp[-B^2/2\kappa t]$, we find
$P(R,t)\propto (2\pi\kappa t)^{-d} R^{d-1}\exp[-R^2/2\kappa t]$
satisfying the Fokker-Planck equation $\partial P/\partial
t=(\kappa/2)\partial^2P/\partial
R^2-(\kappa(d-1)/2)\partial(P/R)/\partial R$, corresponding to the
Langevin equation
\begin{eqnarray}
\frac{dR}{dt}=\kappa\frac{d-1}{2R}+\xi_t. \label{bessel}
\end{eqnarray}
For $d\leq 2$ Brownian motion is recurrent, i.e., the particle
returns to the origin $R=0$, for $d>2$ the particle goes off to
infinity. Setting $\kappa(d-1)/2=2$ we obtain $\kappa=4/(d-1)$,
i.e., $R\rightarrow\infty$ for $\kappa<4$ and $R\rightarrow 0$ for
$\kappa>4$.

Since the tip of the curve $z_t$ is mapped to $a_t$, i.e.,
$h(z_t)=0$, the case $x_t\rightarrow\infty$ for $\kappa<4$
corresponds to a curve never intersecting the real axis, i.e., the
curve is simple. For $\kappa >4$ we have $x_t\rightarrow 0$
corresponding to the case where the tip $z_t$ intersects the real
axis forming a hull. Since the curve is self-similar the
intersections takes place on all scales and eventually the whole
upper half plane is engulfed by the hull.

The marginal value $\kappa = 4$ can also be inferred from a simple
heuristic argument [13]. For small $\kappa$ we can ignore the noise
in Eq. (\ref{sle3})and the particle is repelled according to the
solution $x_t^2\sim 4t$. For large noise we ignore the nonlinear
term and the noise can drive $x_t$ to zero; we have $x_t^2\sim
\kappa t$. The balance is obtained for $\kappa = 4$.

\begin{figure}
\begin{center}
\includegraphics[width=0.7\hsize]
{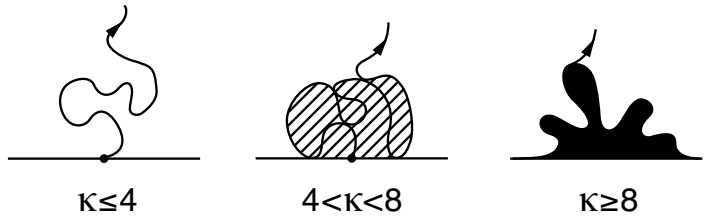}\caption{The figures depict
the phases of SLE. For $\kappa\leq 4$ the SLE trace is a simple
non-intersecting scale invariant random curve from the origin to
infinity with a fractal dimension between $1$ and $3/2$. For
$4<\kappa\leq 8$ the SLE curve is self-intersecting on all scales
and also intersects the real axis on all scales. The curve
together with the enclosed regions, the hull, eventually exhausts
the upper half plane. The scale invariant hull has a fractal
dimension ranging between $3/2$ and $2$. For $\kappa\geq 8$ the
fractal dimension of the hull locks onto $2$ and the scale
invariant hull is dense and plane-filling.} \label{fig8}
\end{center}
\end{figure}

In conclusion, for $\kappa\leq 4$ the curve is simple, for
$\kappa>4$ the curve intersects itself and the real axis infinitely
many times on all scales, eventually the hull swallows the whole
plane. For large $\kappa$ the trace turns out to be plane-filling.
The two cases $\kappa\leq 4$ and $\kappa>4$ are depicted in
Fig.~\ref{fig8}.

\subsection{Fractal Dimension}

The SLE random curves are fractal. An important issue is thus the
determination of the fractal dimension $D$ in terms of the Brownian
strength or SLE parameter $\kappa$. It has been shown that
[45,46,79]

\begin{eqnarray}
D = 1 + \frac{\kappa}{8}~~~~~~~~\text{for}~~0\leq\kappa\leq 8,
\label{frac}
\end{eqnarray}
for $\kappa\geq 8$ the fractal dimension locks onto $2$ and the
SLE curve is plane-filling.

In order to illustrate a typical SLE calculation we follow Cardy
[13] in a heuristic derivation of Eq. (\ref{frac}). In order to
evaluate $D$ and in accordance with its definition [16,27] the
standard procedure is to cover the object with disks of size
$\epsilon$ and follow how the number of disks $N(\epsilon)$ of
size $\epsilon$ scales with $\epsilon$ for small $\epsilon$, i.e.,
$N(\epsilon)\propto\epsilon^{-D}$. However, since the SLE curve is
random the argument has to be rephrased. Alternatively, we
consider a disc of size $\epsilon$ located at a fixed position $z$
and ask for the probability $P(z,\epsilon)$ that the SLE curve
crosses the disk. The number of disks covering an area $A$ is $N_A
= A/\epsilon^2$, i.e., $P\propto \epsilon^{-D}/N_A$, and we infer
$P(z,\epsilon)\propto \epsilon^{2-D}$, where $D$ is the fractal
dimension. Incorporating the Markov property we subject the curve
to an infinitesimal conformal transformation, $h_{\delta t} =
g_{\delta t}-a_{\delta t}$, transforming the point $z$ to $w =
g_{\delta t}(z) - a_{\delta t}$; moreover, all lengths are scaled
by $|h_{\delta t}'(z)|$ [1]. Setting $z=x+iy$ and $w=x'+iy'$ we
obtain expanding Eq. (\ref{sle2}) $x'+iy'=2\delta
t/(x+iy)-\sqrt\kappa\delta B_t$ or $x'=x+2x\delta
t/(x^2+y^2)-\sqrt\kappa\delta B_t$, $y'=y-2y\delta t/(x^2+y^2)$,
$\epsilon'=(1-|h_{\delta t}(z)|)\epsilon$, and $|h_{\delta
t}|=2(x^2-y^2)/(x^2+y^2)^2$. By conformal invariance the
probability measure $P(x,y,\epsilon)$ is unchanged and we infer
\begin{eqnarray}
P(x,y,\epsilon)=\langle P(x',y',\epsilon')\rangle_{\delta B},
\label{prob2}
\end{eqnarray}
where we have averaged over the Brownian motion referring to the
initial part of the curve which has been eliminated by the map
$h_{\delta t}$. Expanding Eq. (\ref{prob2}) to first order in
$\delta t$ and noting that $\langle(\delta B_t)^2\rangle=\delta t$
we arrive at a partial differential equation for $P(x,y,\epsilon)$
\begin{eqnarray}
\left(\frac{2x}{x^2+y^2}\frac{\partial}{\partial x}
-\frac{2y}{x^2+y^2}\frac{\partial}{\partial y}
+\frac{\kappa}{2}\frac{\partial^2}{\partial x^2} -
\frac{2(x^2-y^2)}{(x^2+y^2)^2}\epsilon\frac{\partial}{\partial\epsilon
}\right)P=0.
\end{eqnarray}
Since $P(x,y,\epsilon)\propto\epsilon^{2-D}$ we have
$\epsilon\partial P/\partial\epsilon=(2-D)P$ and the determination
of $D$ is reduced to an eigenvalue problem. By inspection one finds
\begin{eqnarray}
P\propto \epsilon^{1-\kappa/8}y^{(\kappa-8)^2/8\kappa}
(x^2+y^2)^{(\kappa-8)/2\kappa},
\end{eqnarray}
and we identify the fractal dimension $D=1+\kappa/8$ for
$\kappa<8$; for $\kappa>8$ another solution yields $D=2$.

\section{Results and Discussion}
Stochastic Loewner evolution based on Eq. (\ref{sle}) generates
conformally invariant non-crossing random curves in the upper half
plane starting at the origin and going off to infinity. This is the
case of chordal SLE, where the random curve connects two boundary
points (the origin $\bf O$ and infinity $\mathbf\infty$). Another
case is radial SLE for random curves connecting a boundary point and
an interior point in a simply connected domain [6,13,21]. Radial SLE
is governed by another stochastic equation and will not be discussed
here.
%%%%%%%%%%%%%%%%%%%%%%%%%%%%%%%%%%%%%%%%%%%%%%%%%%%%
%%%%%%%%%%%%%%%%%%%%%%%%%%%%%%%%%%%%%%%%%%%%%%%%%%%%
%%%%%%%%%%%%%%%%%%%%%%%%%%%%%%%%%%%%%%%%%%%%%%%%%%%%
\subsection{Phase transitions - Locality - Restriction - Duality}
%%%%%%%%%%%%%%%%%%%%%%%%%%%%%%%%%%%%%%%%%%%%%%%%%%%%
%%%%%%%%%%%%%%%%%%%%%%%%%%%%%%%%%%%%%%%%%%%%%%%%%%%%
%%%%%%%%%%%%%%%%%%%%%%%%%%%%%%%%%%%%%%%%%%%%%%%%%%%%
\subsubsection{Phase Transitions}
\begin{figure}
\begin{center}
\includegraphics[width=0.66\hsize]
{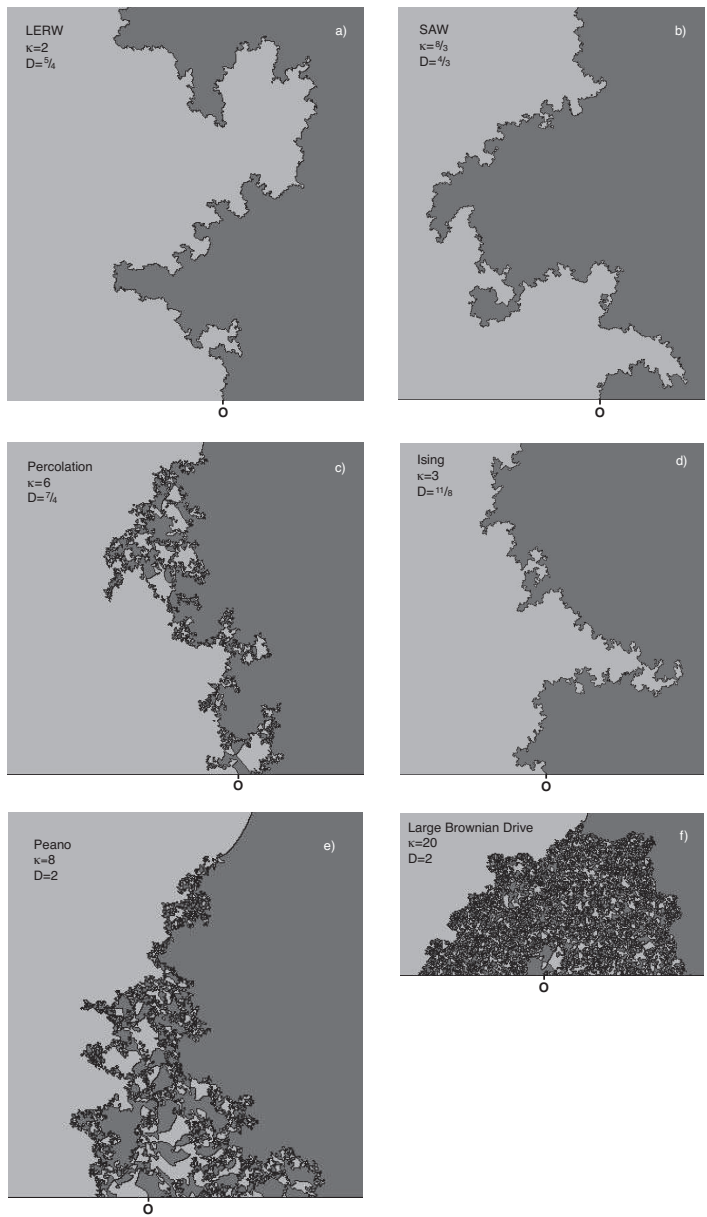}\caption{We depict a
numerical renderings of SLE for a variety of $\kappa$ values. In
a) we show loop erased random walk (LERW) for $\kappa=2$ with
fractal dimension $D=5/4$. In b) we illustrate the case of
self-avoiding random walk (SAW) for $\kappa=8/3$ and fractal
dimension $D=4/3$; both LERW and SAW have $\kappa>4$ and are
simple scale invariant random curves. In c) we depict site
percolation for $\kappa=6$ with fractal dimension $D=7/4$. Since
$\kappa>4$ the percolation case is self-intersecting and duality
implies that the boundary or frontier of the hull is described by
a SLE curve for $\kappa=16/6=8/3$, i.e., the case of SAW. In d) we
show the Ising case for $\kappa=3$ and fractal dimension $D=11/8$.
In e) we depict the limiting case $\kappa=8$ and fractal dimension
$D=2$. The hull is dense and plane-filling. The frontier of the
hull corresponds to the SLE case $\kappa=16/8=2$, i.e., the case
of LERW. The so-called uniform spanning tree (UST) has the same
properties as LERW and the SLE case for $\kappa=8$ can thus be
thought of as a random plane filling Peano curve wrapping around
the UST. Finally, in f) we show the SLE trace and hull for
$\kappa=20$ and $D=2$. Because of the large Brownian excursions
the plane-filling hull is vertically compressed (with permission
from V. Beffara:
http://www.umpa.ens-lyon.fr/$\sim$vbeffara/simu.php).}
\label{fig9}
\end{center}
\end{figure}

%%%%%%%%%%%%%%%%%%%%%%%%%%%%%%%%%%%%%%%%%%%%%%%%%%%%
%%%%%%%%%%%%%%%%%%%%%%%%%%%%%%%%%%%%%%%%%%%%%%%%%%%%
%%%%%%%%%%%%%%%%%%%%%%%%%%%%%%%%%%%%%%%%%%%%%%%%%%%%
SLE exhibits two phase transitions; for $\kappa=4$ and $\kappa=8$.
For $0<\kappa\leq 4$ the random curve is non-intersecting, i.e., a
simple random continuous curve from $\bf O$ to $\infty$. For
$4<\kappa\leq 8$ the curve is self-intersecting on all scales. The
curve together with the excluded regions form a hull which in the
course of time absorbs the upper half plane. For $\kappa$ just
above $4$ the half plane is eventually absorbed but the trace does
not visit all regions, i.e., the hull is not dense. As we approach
$\kappa=8$  the trace becomes more dense and the hull becomes
plane-filling. This is also reflected in the fractal dimension
$D=1+\kappa/8$. For $\kappa> 8$ the hull is plane-filling, i.e.,
$D=2$. As we increase the strength of the Brownian drive further
the excursions of the trace to the right and left in the upper
half plane become more pronounced and the hull becomes vertically
compressed. These results have been obtained by Rohde and Schramm
[79] and Lawler et al. [69]. The various phases of SLE are
depicted in Fig.~\ref{fig8}. In Fig.~\ref{fig9} we have depicted
numerical renderings of SLE traces for various values of $\kappa$
(with permission from V. Beffara,
http://www.umpa.ens-lyon.fr/$\sim$vbeffara/simu.php).

\subsubsection{Locality - Restriction}

In addition to the phase transitions at $\kappa=4$ and $\kappa=8$,
there are special values of $\kappa$ where SLE shows a behavior
characteristic of the scaling limit of specific lattice models: The
locality property for $\kappa=6$ and the restriction property for
$\kappa=8/3$. The issue here is the influence of the boundary on the
SLE trace.

To illustrate the locality property, consider for example the SLE
trace originating at the origin and purporting to describe the
scaling limit of a domain wall in the lattice model. Due to the
long range correlations at the critical point it is intuitively
clear that a deformation of the boundary, e.g., a bulge $\bf L$ on
the real axis to the right of the origin, will influence the trace
and push it to the left. A detailed analysis show that only for
$\kappa=6$ is the trace independent of a change of the boundary,
i.e., the trace does not feel the boundary until it encounters a
boundary point [71,75]. Returning to the lattice models the
locality property for $\kappa=6$ applies specifically to the
percolation case where the interface generated by the exploration
process is governed by a local rule and the model has a geometric
phase transition.

The restriction property is less obvious to visualize but
basically states that the distribution of traces conditioned not
to hit a bulge $\bf L$ on the real axis away from the origin is
the same as the distribution of traces in the domain where $\bf L$
is part of the boundary, i.e., in the domain $\bf H\setminus L$.
Analysis shows that the restriction property only applies in the
case for $\kappa=8/3$. Among the lattice models only the scaling
limit of self-avoiding random walk (SAW), where the measure is
uniform, conforms to the restriction property and thus corresponds
to $\kappa=8/3$ [75].

\subsubsection{Duality}

For $\kappa>4$ the SLE generates a hull of fractal dimension
$D>3/2$. The boundary, external perimeter, or frontier of the hull
is again a simple conformally invariant random curve characterized
by the fractal dimension $\bar D$. Using methods from 2D quantum
gravity Duplantier [53] has proposed the relationship,
\begin{eqnarray}
(D-1)(\bar D-1)={1\over 4}, \label{dual1}
\end{eqnarray}
between the fractal dimension of the hull and its frontier. This
result has been proved by Beffara for $\kappa=6$, i.e., the
percolation case [46]. Inserting in Eq. (\ref{frac}) we obtain for
the corresponding SLE parameter the duality relation
\begin{eqnarray}
\kappa\bar\kappa=16. \label{dual2}
\end{eqnarray}
%
%%%%%%%%%%%%%%%%%%%%%%%%%%%%%%%%%%%%%%%%%%%%%%%%%%%%
%%%%%%%%%%%%%%%%%%%%%%%%%%%%%%%%%%%%%%%%%%%%%%%%%%%%
%%%%%%%%%%%%%%%%%%%%%%%%%%%%%%%%%%%%%%%%%%%%%%%%%%%%
\subsection{Loop Erased Random Walk (LERW)}
%%%%%%%%%%%%%%%%%%%%%%%%%%%%%%%%%%%%%%%%%%%%%%%%%%%%
%%%%%%%%%%%%%%%%%%%%%%%%%%%%%%%%%%%%%%%%%%%%%%%%%%%%
%%%%%%%%%%%%%%%%%%%%%%%%%%%%%%%%%%%%%%%%%%%%%%%%%%%%
Whereas the scaling limit of random walk, i.e., Brownian motion,
does not fall in the SLE category because of self-crossings
rendering Riemann's mapping theorem inapplicable, variations of
Brownian motion are described by SLE.

Loop erased random walk (LERW) where loops are removed along the
way is by construction self-avoiding and was introduced as a
simple model of a self-avoiding random walk. LERW was studied by
Schramm in his pioneering work [82]. LERW has the Markov property
and has been proved to be conformally invariant in the scaling
limit and described by SLE for $\kappa =2$ [69]. According to Eq.
(\ref{frac}) LERW has the fractal dimension $D=5/4$. Also, since
$\kappa<4$ LERW is non-intersecting. A simulation of LERW based on
SLE is shown in Fig.~\ref{fig9}a.
%%%%%%%%%%%%%%%%%%%%%%%%%%%%%%%%%%%%%%%%%%%%%%%%%%%%
%%%%%%%%%%%%%%%%%%%%%%%%%%%%%%%%%%%%%%%%%%%%%%%%%%%%
%%%%%%%%%%%%%%%%%%%%%%%%%%%%%%%%%%%%%%%%%%%%%%%%%%%%
\subsection{Self-avoiding random walk (SAW)}
%%%%%%%%%%%%%%%%%%%%%%%%%%%%%%%%%%%%%%%%%%%%%%%%%%%%
%%%%%%%%%%%%%%%%%%%%%%%%%%%%%%%%%%%%%%%%%%%%%%%%%%%%
%%%%%%%%%%%%%%%%%%%%%%%%%%%%%%%%%%%%%%%%%%%%%%%%%%%%
Self-avoiding random walk (SAW) is a random walk conditioned not to
cross itself. SAW has been used to model polymers in a dilute
solution and has a uniform probability measure. Since SAW satisfies
the restriction property it is conjectured in the scaling limit to
fall in the SLE class with $\kappa = 8/3$ [61,62,63,74], yielding
the fractal dimension $D=4/3$. We note that Flory's mean field
theory [15] for the size $R$ of a polymer composed of $N$ links
(monomers) scales like $R\sim N^\nu$, where $\nu=3/(2+d)$ for $d\leq
4$. By a box covering we infer $N\sim R^D$ where $D$ is the fractal
dimension, i.e., $D=(2+d)/3$. In $d=2$ we obtain $D=4/3$ in
accordance with the SLE result. SLE induced SAW in the scaling limit
is shown in Fig.~\ref{fig9}b.
%%%%%%%%%%%%%%%%%%%%%%%%%%%%%%%%%%%%%%%%%%%%%%%%%%%%
%%%%%%%%%%%%%%%%%%%%%%%%%%%%%%%%%%%%%%%%%%%%%%%%%%%%
%%%%%%%%%%%%%%%%%%%%%%%%%%%%%%%%%%%%%%%%%%%%%%%%%%%%
\subsection{Percolation}
%%%%%%%%%%%%%%%%%%%%%%%%%%%%%%%%%%%%%%%%%%%%%%%%%%%%
%%%%%%%%%%%%%%%%%%%%%%%%%%%%%%%%%%%%%%%%%%%%%%%%%%%%
%%%%%%%%%%%%%%%%%%%%%%%%%%%%%%%%%%%%%%%%%%%%%%%%%%%%
The scaling limit of site percolation was conjectured by Schramm
[82] to fall in the SLE class for $\kappa = 6$. Subsequently, the
scaling limit of site percolation on a triangular lattice has been
proven by Smirnov [83,84]. Percolation exhibits a geometrical
phase transition. In the exploration process defining a critical
interface the rule for propagation is entirely local. The lack of
stiffness as for example in the Ising case to be discussed below
results in a strongly meandering path winding back and in the
scaling limit intersecting earlier part of the path. Since
$\kappa>4$ the path together with the enclosed part, i.e., the
hull, eliminates the whole plane in the course of time. As
discussed above the locality property is specific to percolation
and yields $\kappa=6$. The fractal dimension of the percolation
interface is according to Eq. (\ref{frac}) $D=7/4$. We note that
$D$ is close to $2$, i.e., the percolation interface nearly covers
the plane densely. A series of new results and a proof of Cardy's
conjectured formula for the crossing probability have appeared; we
refer to [6,13,21] for details. Using the duality relation
(\ref{dual2}) the frontier of the percolation hull is a simple SLE
curve for $\kappa=8/3$, corresponding to SAW. In Fig.~\ref{fig9}c
we have depicted a SLE generated percolation interface.
%%%%%%%%%%%%%%%%%%%%%%%%%%%%%%%%%%%%%%%%%%%%%%%%%%%%
%%%%%%%%%%%%%%%%%%%%%%%%%%%%%%%%%%%%%%%%%%%%%%%%%%%%
%%%%%%%%%%%%%%%%%%%%%%%%%%%%%%%%%%%%%%%%%%%%%%%%%%%%
\subsection{Ising Model - O(n) Models}
%%%%%%%%%%%%%%%%%%%%%%%%%%%%%%%%%%%%%%%%%%%%%%%%%%%%
%%%%%%%%%%%%%%%%%%%%%%%%%%%%%%%%%%%%%%%%%%%%%%%%%%%%
%%%%%%%%%%%%%%%%%%%%%%%%%%%%%%%%%%%%%%%%%%%%%%%%%%%%
The Ising model in Eq. (\ref{ham}) is a special case of the O(n)
model defined by the Hamiltonian
\begin{eqnarray}
H=-J\sum_{\langle ij\rangle}\vec\sigma_i\vec\sigma_j, \label{ham2}
\end{eqnarray}
where $\vec\sigma_i=(\sigma_1,\cdots\sigma_n)$ is an n-component
unit vector associated with the site $i$. For $n=1$ we recover the
Ising model, $n=2$ is the XY-model [14], and $n=3$ the Heisenberg
model.

By means of the Fortuin-Kasteleyn (FK) transformation based on a
high temperature expansion the configurations of the O(n) model can
be described by clusters or graphs on a dual lattice [55]. The
crossing domain wall in Fig.~\ref{fig1} is thus a special case of a
FK graph if we interpret the representation as a triangular Ising
model. It has been conjectured that $n$ is related to the SLE
parameter $\kappa$ by
\begin{eqnarray}
n=-2\cos(4\pi/\kappa)~~\text{for}~~8/3\leq\kappa\leq 4.
\label{on-model}
\end{eqnarray}
In the Ising case $n=1$ and we have  $\kappa=3$ yielding the fractal
dimension $D=11/8$ for the Ising domain wall [86]. Since $\kappa<4$
the Ising domain wall is non-intersecting. Unlike the percolation
case, the Ising interface is stiffer due to the interaction. We also
note that the scaling limit of spin cluster boundaries in the Ising
model recently has been proven to correspond to SLE for $\kappa = 3$
[85]. The interface is shown in Fig.~\ref{fig9}d.
%%%%%%%%%%%%%%%%%%%%%%%%%%%%%%%%%%%%%%%%%%%%%%%%%%%%
%%%%%%%%%%%%%%%%%%%%%%%%%%%%%%%%%%%%%%%%%%%%%%%%%%%%
%%%%%%%%%%%%%%%%%%%%%%%%%%%%%%%%%%%%%%%%%%%%%%%%%%%%
\subsection{SLE - Conformal Field Theory}
%%%%%%%%%%%%%%%%%%%%%%%%%%%%%%%%%%%%%%%%%%%%%%%%%%%%
%%%%%%%%%%%%%%%%%%%%%%%%%%%%%%%%%%%%%%%%%%%%%%%%%%%%
%%%%%%%%%%%%%%%%%%%%%%%%%%%%%%%%%%%%%%%%%%%%%%%%%%%%

Whereas conformal field theory (CFT) is based on the concept of a
local field $\phi(r)$ and its correlations and therefore only
access the underlying geometry indirectly through field
correlations, SLE directly produces conformally invariant
geometrical objects. A major issue is therefore the connection
between CFT and SLE [39,40,51]. In CFT the central charge $c$
plays an important role in delimiting the universality classes of
the variety of lattice models yielding conformal field theories in
the scaling limit. Percolation thus corresponds to the central
charge $c=0$, whereas the Ising model is associated with the
central charge $c=1/2$. It has been conjectured that the
connection between the SLE parameter $\kappa$ and the central
charge $c$ is given by
\begin{eqnarray}
c=\frac{(6-\kappa)(3\kappa-8)}{2\kappa}=1-6\frac{(\kappa-4)^2}{4\kappa}.
\label{cft}
\end{eqnarray}
We note that $c<1$ and, moreover, invariant under the duality
tranformation $\kappa\rightarrow 16/\kappa$.
%%%%%%%%%%%%%%%%%%%%%%%%%%%%%%%%%%%%%%%%%%%%%%%%%%%%
%%%%%%%%%%%%%%%%%%%%%%%%%%%%%%%%%%%%%%%%%%%%%%%%%%%%
%%%%%%%%%%%%%%%%%%%%%%%%%%%%%%%%%%%%%%%%%%%%%%%%%%%%
\subsection{SLE - 2D turbulence}
%%%%%%%%%%%%%%%%%%%%%%%%%%%%%%%%%%%%%%%%%%%%%%%%%%%%
%%%%%%%%%%%%%%%%%%%%%%%%%%%%%%%%%%%%%%%%%%%%%%%%%%%%
%%%%%%%%%%%%%%%%%%%%%%%%%%%%%%%%%%%%%%%%%%%%%%%%%%%%
There is an interesting application of SLE ideas in the context of
2D turbulence. The issue here is to analyze conformal invariance by
comparing the statistical properties of geometrical shapes like
domain walls with SLE traces with the view of determining the SLE
parameter $\kappa$ and the corresponding universality class.

In 3D turbulence is governed by the incompressible Navies-Stokes
equation for the velocity field. Since the viscosity is only
effective at small length scales 3D turbulence is characterized by a
cascade of kinetic energy $(1/2)v^2$ from large scales (driving
scale) to small scales (dissipation scale). In the inertial regime
the energy spectrum $E(k)$ ($k$ is the wavenumber) is characterized
by the celebrated Kolmogorov $5/3$ law [61], $E(k)\propto k^{-5/3}$,
indicating an underlying scale invariance in turbulence.

In 2D the cascade picture is different. Since both kinetic energy
and squared vorticity (enstrophy) are conserved in the absence of
dissipation and forcing, two cascades coexist [67,68]. A direct
cascade to small scales for the squared vorticity
$\omega^2=(\nabla\times v)^2$ with scaling exponent $-3$ and an
inverse cascade to larger scales for the kinetic energy $(1/2)v^2$
with Kolmogoroff scaling exponent $-5/3$. The system is thus
characterized by a fine scale vorticity structure together with a
large scale velocity structure. Moreover, we can assume that the
vorticity structure is equipartitioned, i.e, in equilibrium.

In order to investigate whether the scale invariance of the small
scale vorticity structure can be extended to conformal invariance
Bernard et al [48] have considered the statistics of the boundaries
of vorticity clusters. By comparing the zero-vorticity isolines with
SLE traces they find that cluster boundaries fall in the
universality class corresponding to $\kappa = 6$, i.e., the case of
percolation. Since 2D turbulence is a driven nonequilibrium system,
this observation is very intriguing in particular since the
correlations between vortices are long-ranged. A similar analysis
[49] of the isolines in the inverse cascade in surface
quasigeostrophic turbulence corresponds to $\kappa = 4$, i.e., from
Eq. (\ref{on-model}) the domain walls in the equilibrium XY model
for $n=2$. For comments on the application of SLE in turbulence we
refer to Cardy [52].
%%%%%%%%%%%%%%%%%%%%%%%%%%%%%%%%%%%%%%%%%%%%%%%%%%%%
%%%%%%%%%%%%%%%%%%%%%%%%%%%%%%%%%%%%%%%%%%%%%%%%%%%%
%%%%%%%%%%%%%%%%%%%%%%%%%%%%%%%%%%%%%%%%%%%%%%%%%%%%
\subsection{SLE - 2D spin glass}
%%%%%%%%%%%%%%%%%%%%%%%%%%%%%%%%%%%%%%%%%%%%%%%%%%%%
%%%%%%%%%%%%%%%%%%%%%%%%%%%%%%%%%%%%%%%%%%%%%%%%%%%%
%%%%%%%%%%%%%%%%%%%%%%%%%%%%%%%%%%%%%%%%%%%%%%%%%%%%
It is a standing issue whether conformal field theory can be applied
to disordered systems, in particular systems with quenched disorder.
In recent work Amoruso et al. [38] and Bernard et al. [47] have
considered zero temperature domain walls in the Ising spin glass
[17]; see also [54]. The Ising spin glass is an equilibrium system
with quenched disorder. The system is described by the Hamiltonian
$H=-\sum_{\langle ij\rangle}J_{ij}\sigma_i\sigma_j$, where the
random exchange constants $J_{ij}$ are picked from a Gaussian
distribution with zero mean. The glass transition is at $T=0$ and
the system has a two-fold degenerate ground state. Inducing a scale
invariant domain wall between the two ground states and comparing
with an SLE trace, it is found that both the Markov and conformal
properties are obeyed and that the universality class corresponds to
$\kappa\approx 2.3$.
%%%%%%%%%%%%%%%%%%%%%%%%%%%%%%%%%%%%%%%%%%%%%%%%%%%%
%%%%%%%%%%%%%%%%%%%%%%%%%%%%%%%%%%%%%%%%%%%%%%%%%%%%
%%%%%%%%%%%%%%%%%%%%%%%%%%%%%%%%%%%%%%%%%%%%%%%%%%%%
\subsection{Further remarks}
%%%%%%%%%%%%%%%%%%%%%%%%%%%%%%%%%%%%%%%%%%%%%%%%%%%%
%%%%%%%%%%%%%%%%%%%%%%%%%%%%%%%%%%%%%%%%%%%%%%%%%%%%
%%%%%%%%%%%%%%%%%%%%%%%%%%%%%%%%%%%%%%%%%%%%%%%%%%%%
In this discussion have left out several topics which have played
an important role in the development and applications of SLE. We
mention some of them below.

There is an interesting connection between LERW and the so-called
uniform spanning tree (UST) [6,21,69,82]. A spanning tree is a
collection of vertices and edges which form a tree, i.e., without
loops or cycles. A uniform spanning tree is a random spanning tree
picked among all possible spanning trees with equal probability.
Consider the unique path between two vertices on a UST. Since the
path lives on a tree it is by construction non-crossing and it turns
out that it has the same distribution as LERW.  The winding random
curve enclosing the UST can be visualized as a random plane-filling
Peano curve. In the scaling limit the Peano curve is described by
SLE for $\kappa=8$ with fractal dimension $D=2$.

The q-state Potts model [37] constitutes a generalization of the
Ising model; here the lattice variable takes $q$ values. The model
is defined by the Hamiltonian $H=-J\sum_{\langle
ij\rangle}\delta_{\sigma_i\sigma_j}$, where $\sigma_i=1,\cdots q$;
the Ising model obtains for $q=2$. Applying the high temperature
FK representation the configurations can be represented by loops
and domain walls. From considerations involving the fractal
dimension [81] it has been conjectured that domain walls in the
scaling limit of the Potts model fall in the SLE category for
$q=2+2\cos(8\pi/\kappa)$, where $4\leq\kappa\leq 8$. For $q=2$ we
recover the Ising case for $\kappa=3$. In the limit $q\rightarrow
0$, the graph representation is equivalent to the uniform spanning
tree described by SLE for $\kappa=8$. For a numerical study of the
three-state Potts model and its relation to SLE consult [56].

Standard SLE is driven by 1D Brownian motion producing a fractal
curve. Ruskin et al. [80] have considered the case of adding a
stable L\'evy process with shape parameter $\alpha$ to the Brownian
motion. Backing their analysis with numerics they find that the SLE
trace branches and exhibit a 'phase transitions' related to
self-intersections.

2D Brownian motion, the scaling limit of 2D random walk, is an
incredibly complex fractal coil owing to the self-crossings on all
scales. Although 2D Brownian motion because of self-crossing
itself falls outside the SLE scheme, the outer frontier or
perimeter of 2D random walk is a non-crossing and non-intersecting
fractal curve which can be accessed by SLE. Verifying an earlier
conjecture by Mandelbrot [27] it has been proven using SLE
techniques [70] that the fractal dimension of the Brownian
perimeter is $D=4/3$, i.e, the same as the fractal dimension of
self-avoiding random walk and the external perimeter of the
percolation hull. Other characteristics of Brownian motion such as
intersection exponents have also been obtained [71,72,73]; see
also [77].

\section{Future directions}

Stochastic Loewner evolution represents a major step in our
understanding of fractal shapes in the 2D continuum limit. By
combining the Markov property (stationarity) with conformal
invariance SLE provides a minimal scheme for the generation of a
one-parameter family of fractal curves. The SLE scheme also
provides calculational tools which have led to a host of new
results. SLE is a developing field and we can on the mathematical
front anticipate progress and proofs of some yet unproven scaling
limits, e.g., the scaling limit of the FK representation of the
Potts model and the scaling limit of SAW.

On the more physical front many issues also remain open. First there
is the fundamental issue of the connection between the hugely
successful but non-rigorous CFT and SLE. Here progress is already
under way. In a series of papers Bauer and Bernard
[39,40,41,42,43,44] have shown how SLE results can be derived using
CFT methods. Cardy [51] have considered a multiple SLE process and
the connection to Dyson's Brownian process and random matrix theory.
The analysis of the CFT-SLE connection still remains to be analyzed
further.

An obvious limitation of SLE is that it only addresses critical
domain walls and not the full configuration of clusters and loops in
for example the FK representation of the Potts model. In the case of
critical percolation this problem has been addressed by Camia and
Newman [50]. Another issue is how to provide SLE insight into spin
correlations in the Potts or O(n) models.

In the original formulation of SLE the Markov and conformal
properties essentially requires a Brownian drive. It is clearly of
interest to investigate the properties of random curves generated by
other random drives. Such a program has been initiated by  Ruskin et
al. [80] who considered adding a L\'evy drive to the Brownian drive;
see also work by Kennedy [64,65].

Since the SLE trace lives in the infinite upper half plane the
whole issue of finite size effects remain open. In ordinary
critical phenomena the concept of a Kadanoff block construction
and the diverging correlation length near the transition lead to a
theory of finite size scaling and corrections to scaling which can
be accessed numerically. It is an open problem how to develop a
similar scheme for SLE.

In statistical physics it is customary and natural to associate a
free energy to a domain wall and an interaction energy associated
with several domain walls. These free energy considerations are
entirely absent in the SLE framework which is based on conformal
transformations. A major issue is thus: Where is the free energy
in all this and how do we reintroduce and make use of ordinary
physical considerations and estimates [78].

\newpage
\section{Bibliography}
%%%%%%%%%%%%%%%%%%%%%%%%%%%%%%%%%%
%%%%%%%%%%%%%%%%%%%%%%%%%%%%%%%%%%
%%%%%%%%%%%%%%%%%%%%%%%%%%%%%%%%%%
%%%%%%%%%%%%%%%%%%%%%%%%%%%%%%%%%%
{\bf Books and Reviews}
%%%%%%%%%%%%%%%%%%%%%%%%%%%%%%%%%%
%%%%%%%%%%%%%%%%%%%%%%%%%%%%%%%%%%
%%%%%%%%%%%%%%%%%%%%%%%%%%%%%%%%%%
%%%%%%%%%%%%%%%%%%%%%%%%%%%%%%%%%%

[1] Ahlfors LV (1966) Complex analysis: an introduction to the
theory of analytical functions of one complex variable. McGraw-Hill,
New York

[2] Ahlfors LV (1973) Conformal invariance: topics in geometric
function theory. McGraw-Hill, New York

[3] Ash RB, Dol\'eans CA (2000) Probability $\&$ Measure Theory.
Academic Press, San Diego

[4] Bak P (1999) How Nature Works: The Science of Self-Organized
Criticality. Springer, New York

[5] Bauer M, Bernard D (2004)  Loewner Chains.
arXiv:cond-mat/0412372

[6] Bauer M, Bernard D (2006) 2D growth processes: SLE and Loewner
chains. Physics Reports,432:115-221

[7] Baxter RJ (1982) Exactly solved models in statistical
mechanics. Academic Press, London

[8] Binney JJ, Dowrick NJ, Fisher AJ, Newman MEJ (1992) The Theory
of Critical Phenomena. Clarendon Press, Oxford

[9] Cardy J (1987) Conformal invariance. In Phase Transitions and
Critical Phenomena, vol 11, eds. Domb C and Lebowitz JL, Academic
Press, London

[10] Cardy J (1993) Conformal field theory comes of age. Physics
World, June, 29-33

[11] Cardy J (1996) Scaling an Renormalization in Statistical
Physics. Cambridge University Press, Cambridge

[12] Cardy J (2002) Conformal Invariance in Percolation,
Self-Avoiding Walks and Related Problems. Plenary talk given at
TH-2002, Paris; arXiv:cond-mat/0209638

[13] Cardy J (2005) SLE for theoretical physicists. Ann. Phys.
318:81-118; arXiv:cond-mat/0503313

[14] Chaikin PM, Lubensky TC (1995) Principles of Condensed Matter
Physics. Cambridge University Press, Cambridge

[15] de Gennes PG (1985) Scaling concepts in polymer physics,
Cornell University Press, Ithaca

[16] Feder J (1988) Fractals (Physics of Solids and Liquids).
Springer, New York

[17] Fischer KH, Hertz JA (1991) Spin Glasses. Cambridge University
Press, Cambridge

[18] Gardiner CW (1997) Handbook of Stochastic Methods.
Springer-Verlag, New York

[19] Gong S (1999) The Bieberbach Conjecture. R.I. American 19.
Mathematical Society, International Press, Providence

[20] Jensen HJ (2000) Self-Organized Criticality: Emergent Complex
Behavior in Physical and Biological Systems. Cambridge University
Press, Cambridge

[21] Kager W, Nienhuis B (2004) A guide to Stochastic Loewner
evolution and its application. J. Stat. Phys. 115:1149-1229

[22] Kauffman SA (1996) At Home in the Universe: The Search for the
Laws of Self-Organization and Complexity. Oxford University Press,
Oxford

[23] Landau LD, Lifshitz EM (1959) Theory of Elasticity. Pergamon
Press, Oxford

[24] Lawler GF (2005) Conformally invariant processes in the plane.
Mathematical Surveys, 114, AMS, Providence, RI

[25] Lawler GF (2004) ICTP Lecture Notes Series,

[26] Ma S-K. (1976) Modern theory of critical phenomena. Frontiers
in Physics, vol 46, Benjamin, Reading

[27] Mandelbrot B (1987) The Fractal Geometry of Nature. W.H.
Freeman \& Company

[28] Nicolis G (1989) Exploring Complexity: An Introduction. W.H.
Freeman \& Company

[29] Nienhuis B (1987) Coulomb gas formulation of two-dimensional
phase transitions. In Phase Transitions and Critical Phenomena,
vol 11, eds. Domb C and Lebowitz JL, Academic Press, London

[30] Pfeuty P, Toulouse G (1977) Introduction to the Renormalization
Group and to Critical Phenomena. Wiley, New York

[31] Reichl LE (1998) A Modern Course in Statistical Physics.
Wiley, New York

[32] Stanley HE (1987) Introduction to Phase Transitions and
Critical Phenomena. Oxford University Press, Oxford

[33] Stauffer D, Aharony A (1994) Introduction To Percolation
Theory. CRC

[34] Strogatz S (2003) Sync: The Emerging Science of Spontaneous
Order. Hyperion

[35] Werner W (2004) Random planar curves and Schramm-Loewner
evolutions. Springer Lecture Notes in Mathematics 1840:107-195;
arXiv: math.PR/0303354

[36] Wilson KG, Kogut J (1974) The renormalization group and the
$\epsilon$ expansion. Physics Reports,12:75-199

[37] Wu FY (1982) The Potts model. Rev. Mod. Phys. 54:235-268

%%%%%%%%%%%%%%%%%%%%%%%%%%%%%%%%%%
%%%%%%%%%%%%%%%%%%%%%%%%%%%%%%%%%%
%%%%%%%%%%%%%%%%%%%%%%%%%%%%%%%%%%
%%%%%%%%%%%%%%%%%%%%%%%%%%%%%%%%%%
{\bf Primary literature}
%%%%%%%%%%%%%%%%%%%%%%%%%%%%%%%%%%
%%%%%%%%%%%%%%%%%%%%%%%%%%%%%%%%%%
%%%%%%%%%%%%%%%%%%%%%%%%%%%%%%%%%%
%%%%%%%%%%%%%%%%%%%%%%%%%%%%%%%%%%

[38] Amoruso C, Hartmann AK, Hastings MB, Moore MA (2006)
Conformal Invariance and Stochastic Loewner Evolution Processes in
Two-Dimensional Ising Spin Glasses. Phys. Rev. Lett. 97:267202(4);
arXiv:cond-mat/0601711

[39] Bauer M, Bernard D (2002) $\text{SLE}_\kappa$ growth processes
and conformal field theory. Phys. Lett. B 543:135-138; arXiv:
math.PR/0206028

[40] Bauer M, Bernard D (2003) Conformal field theories of
Stochastic Loewner evolutions. Comm. Math. Phys. 239:493-521;
arXiv: hep-th/0210015

[41] Bauer M, Bernard D (2003) SLE martingales and the Viasoro
algebra. Phys. Lett. B 557: 309-316; arXiv: hep-th/0301064

[42] Bauer M, Bernard D (2004) Conformal transformations and the SLE
partition function martingale. Annales Henri Poincare 5:289-326;
arXiv:math-ph/0305061

[43] Bauer M, Bernard D (2004) CFTs of SLEs: the radial case.
Phys.Lett. B 583:324-330;  arXiv:math-ph/0310032

[44] Bauer M, Bernard D (2004) SLE, CFT and zig-zag probabilities.
Proceedings of the conference `Conformal Invariance and Random
Spatial Processes', Edinburgh, July 2003;  arXiv:math-ph/0401019

[45] Beffara V (2002) The dimension of SLE curves;
arXiv:math.PR/0211322

[46] Beffara V (2003) Hausdorff dimensions for SLE$_6$. Ann. Probab.
32:2606-2629; arXiv:math.PR/0204208

[47] Bernard D, Le Doussal P., Middleton AA (2006) Are Domain Walls
in 2D Spin Glasses described by Stochastic Loewner Evolutions.
arXiv:cond-mat/0611433

[48] Bernard D, Boffetta G, Celani A, Falkovich G (2006) Conformal
invariance in two-dimensional turbulence. Nature Physics, 2:124-128

[49] Bernard D, Boffetta G, Celani A, Falkovich G (2007) Inverse
Turbulent Cascades and Conformally Invariant Curves. Phys. Rev.
Lett. 98:024501(4); arXiv: nlin.CD/0602017

[50] Camia F, Newman CM (2003)   Continuum nonsimple loops and 2D
critical percolation. arXiv: math.PR/0308122

[51] Cardy J (2003) Stochastic Loewner evolution and Dyson's
circular ensembles. J. Phys. A 36: L379-L408; arXiv: math-ph/0301039

[52] Cardy J (2006) The power of two dimensions. Nature Physics,
2:67-68

[53] Duplantier B (2000) Conformally Invariant Fractals and
Potential Theory. Phys. Rev. Lett. 84:1363-1367;
arXiv:cond-mat/9908314

[54] Fisch R (2007) Comment on "Conformal invariance and stochastic
Loewner evolution processes in two-dimensional Ising spin glasses.
arXiv:0705.0046

[55] Fortuin CM, Kasteleyn PW (1972) On the random cluster model.
Physica 57:536-564

[56] Gamsa A, Cardy J (2007) SLE in the three-state Potts model - a
numerical study;  arXiv:0705.1510

[57] Gruzberg IA, Kadanoff LP (2004) The Loewner Equation: Maps and
Shapes. J. Stat. Phys. 114:1183-1198; arXiv:cond-mat/0309292

[58] Kadanoff LP (1966) Scaling laws for Ising models near $T_c$.
Physics 2:263-271

[59] Kadanoff LP, Berkenbusch MK (2004) Trace for the Loewner
equation with singular forcing. Nonlinearity 17:R41-R54;
arXiv:cond-mat/0402142

[60] Kager W, Nienhuis B, Kadanoff LP (2004) Exact Solutions for
Loewner Evolutions. J. Stat. Phys. 115:805-822

[61] Kennedy T (2002) Monte Carlo Tests of Stochastic Loerwner
Evolution Predictions for the 2D self-avoiding walk. Phys. Rev.
Lett. 88:130601(4); arXiv: math.PR/0112246

[62] Kennedy T (2004) Conformal invariance and Stochastic Loewner
evolution predictions for the 2D Self-Avoiding walk - Monte Carlo
tests. J. Stat. Phys. 114:51-78; arXiv: math.PR/0207231

[63] Kennedy T (2005) Monte Carlo comparisons of the self-avoiding
walk and SLE as parameterized curves. arXiv:math.PR/0510604v1

[64] Kennedy T (2006) The length of an SLE - Monte Carlo studies.
arXiv:math.PR/0612609v1

[65] Kennedy T (2007) Computing the Loewner driving process of
random curves in the half plane. arXiv:math.PR/0702071v1

[66] Kolmogorov AN (1941) Dissipation of energy in the locally
isotropic turbulence. Dokl. Akad. Nauk SSSR 30:9-13; reprinted in
Proc. R. Soc. Lond. A 434:9-13 (1991)

[67] Kraichnan RH (1967) Inertial ranges in two-dimensional
turbulence. Phys. Fluids. 10:1417-1423

[68] Kraichnan RH, Montgomery D (1980) Two-dimensional turbulence.
Rep. Prog. Phys. 43:567-619

[69] Lawler GF, Schramm O, Werner W (2001) Conformal invariance of
planar loop-erased random walks and uniform spanning trees. Ann.
Prob. 32:939-995; arXiv: math.PR/0112234

[70] Lawler GF, Schramm O, Werner W (2001) The dimension of the
planar Brownian frontier is $4/3$. Math. Res. Lett. 8:401-411;
arXiv: math.PR/00010165

[71] Lawler GF, Schramm O, Werner W (2001) Values of Brownian
intersections exponents I: half plane exponents. Acta Mathematica
187:237-273; arXiv: math.PR/9911084

[72] Lawler GF, Schramm O, Werner W (2001) Values of Brownian
intersections exponents II: plane exponents. Acta Mathematica
187:275-308; arXiv: math.PR/0003156

[73] Lawler GF, Schramm O, Werner W (2002) Values of Brownian
intersections exponents III: two-sided exponents. Ann. Inst. Henri
poincare 38:109-123; arXiv: math.PR/0005294

[74] Lawler GF, Schramm O, Werner W (2002) On the scaling limit of
planar self-avoiding walk. Fractal geometry and application, A
jubilee of Benoit Mandelbrot, Part 2, 339-364, Proc. Sympos. Pure
Math., 72, Part 2, Amer. Math. Soc., Providence, RI, 2004; arXiv:
math.PR/0204277

[75] Lawler GF, Schramm O, Werner W (2003) Conformal restriction:
The chordal case. J. Amer. Math. Soc. 16:917-955; arXiv:
math.PS/0209343

[76] L\"owner K (Loewner C) (1923) Untersuchungen \"uber schlichte
konforme Abbildungen des Einheitskreises. I. Math. Ann. 89:103-121

[77] Mackenzie D (2000) Taking the Measure of the Wildest Dance on
Earth. Science 290:1883-1884

[78] Moore M (2007) private communication

[79] Rohde S, Schramm O (2001) Basic properties of SLE. Ann. Math.,
vol 161:879-920; arXiv: mathPR/0106036

[80] Rushkin I, Oikonomou P, Kadanoff LP, Gruzberg IA (2006)
Stochastic Loewner evolution driven by Levy processes. J. Stat.
Mech. (2006) P01001(21); arXiv:cond-mat/0509187

[81] Saleur H, Duplantier B (1987) Exact determination of the
percolation hull exponent in two dimensions. Phys. Rev. Lett.
58:2325-2328

[82] Schramm O (2000) Scaling limit of loop-erased random walks and
uniform spanning trees. Israel J. Math. 118:221-288;
arXiv:math.PR/9904022

[83] Smirnov S (2001) Critical percolation in the plane: conformal
invariance, Cardy's formula, scaling limits. C. R. Acad. Sci. Paris
Ser. I Math, 333(3):239-244

[84] Smirnov S, Werner W (2001) Critical exponents for
two-dimensional percolation. Math. Res. Lett. 8:729-744

[85] Smirnov S (2006) Towards conformal invariance of 2D lattice
models. Proceedings of the International Congress of Mathematicians
(Madrid, August 22-30, 2006), European Mathematical Society
2:1421-1451

[86] Vanderzande C, Stella AL (1989) Bulk, surface and hull fractal
dimension of critical Ising clusters in $d=2$. J. Phys. A: Math.
Gen. 22:L445-L451

\newpage

\end{document}